\documentclass[pdflatex,sn-mathphys-num,iicol]{sn-jnl}
\usepackage[T1]{fontenc}
\usepackage[expansion=false]{microtype}
\usepackage{amsmath,amssymb}
\usepackage{graphicx}
\usepackage{booktabs,array}
\usepackage{xcolor}
\usepackage{float}
\usepackage{placeins}
\usepackage{bookmark}
\hypersetup{hidelinks,hypertexnames=false,pdftitle={Fast prediction of electron-beam parameters using transfer maps and machine learning in a digital model of a MeV gamma-gamma collider}}
\graphicspath{{figures/}}
\usepackage{stfloats}
\begin{document}
\title[Fast electron-beam prediction for a MeV $\gamma\gamma$ collider]{Fast prediction of electron-beam parameters using transfer maps and machine learning in a digital model of a MeV $\gamma\gamma$ collider}
\author[1]{\fnm{Jiawei} \sur{Zhang}}\email{zhangjw267@mail2.sysu.edu.cn}
\author[1]{\fnm{Yongsheng} \sur{Huang}}\email{huangysh59@mail.sysu.edu.cn}
\author[1]{\fnm{Yuancun} \sur{Nie}}\email{nieyc5@mail.sysu.edu.cn}
\author[2]{\fnm{Qiang} \sur{Gu}}\email{guq@sari.ac.cn}
\author[3]{\fnm{Yuan} \sur{Chen}}\email{chenyuan79@whu.edu.cn}
\author*[1]{\fnm{Yuanjie} \sur{Bi}}\email{biyj@mail.sysu.edu.cn}
\affil[1]{\orgdiv{School of Science}, \orgname{Shenzhen Campus of Sun Yat-sen University}, \orgaddress{\city{Shenzhen}, \postcode{518107}, \country{China}}}
\affil[2]{\orgname{Shanghai Advanced Research Institute, Chinese Academy of Sciences}, \orgaddress{\city{Shanghai}, \country{China}}}
\affil[3]{\orgname{Wuhan University}, \orgaddress{\city{Wuhan}, \country{China}}}

% ===================================================================== ABSTRACT
\abstract{Tuning the electron beam of a MeV-scale $\gamma\gamma$ collider, first in the design and later in a digital twin of the machine, requires the beam parameters at the interaction point (IP) for many injector states. Start-to-end particle tracking is too slow for this and needs the full particle distribution, which an operating machine cannot measure. We address this problem for the 200-MeV transport line of the collider with two fast predictors. For design, second-order transfer maps from the transport entrance to the IP are applied to every particle. For a digital twin, an extremely-randomized-trees (ExtraTrees) model predicts the IP parameters from the means and rms widths of the beam at the transport entrance. On 89 held-out start-to-end simulations, the map predictor reproduces Elegant tracking for the mean energy, momentum spread, bunch length, transverse rms sizes and normalized emittances with coefficients of determination ($R^2$) of 0.97--1.00 and is ten times faster. The ExtraTrees model reaches an average $R^2$ of 0.88, compared with 0.58 when the injector settings are used as input, and performs similarly with eleven inputs that can be measured or reconstructed. Under a 2\% input perturbation, the average $R^2$ of the six parameters other than the mean energy decreases only from 0.917 to 0.895, whereas the mean energy requires an accurate momentum calibration. The two predictors provide the IP-beam evaluation for the collider design and the prediction layer of a future digital twin.}

\keywords{gamma--gamma collider, beam transport, transfer map, digital twin, machine learning, virtual diagnostics}
\maketitle

% ================================================================= INTRODUCTION
\section{Introduction}\label{sec:intro}
A $\gamma\gamma$ collider converts two electron beams into high-energy photon beams by inverse Compton scattering of laser light at a conversion point (CP) placed shortly before the interaction point (IP)~\cite{takahashi2018}. In a MeV-scale design the photon luminosity and spectrum depend directly on the electron bunch delivered to the CP and IP: its mean energy and momentum spread set the scattered-photon energy and bandwidth, and its transverse size, emittance and bunch length set the overlap with the laser and with the opposing beam. The background from the inverse Compton process has been estimated for such a collider~\cite{zhou2026}, and magnetic-compression and velocity-bunching injectors have been compared for a 200-MeV, 1-nC electron driver with a common double-DBA transport and final-focus system~\cite{zhang2026epjst}.

Such a facility is developed in stages that place different demands on beam modelling. In the \emph{design} stage, start-to-end simulations follow the full particle distribution from the gun to the IP, and candidate injector states are compared by tracking each of them through the downstream line. In a \emph{digital twin}, a fast virtual model of the machine is kept consistent with its configuration and is used to evaluate beam states quickly~\cite{edelen2018,mayes2021,brynes2026}. Finally, when the twin is \emph{connected to the real accelerator}, beam diagnostics supply its inputs and its predictions guide beam tuning. Brynes \emph{et al.} recently presented an architecture that generates digital twins automatically from lattice descriptions, integrates them with the control system and uses them for virtual commissioning and optimization; they identify validation against measured beams as a prerequisite for bidirectional hardware control~\cite{brynes2026}. Each stage needs the beam parameters at the IP, but from different information: the design stage has the complete simulated distribution, whereas an operating machine provides only a few measured or reconstructed beam moments.

Fast surrogate models can replace repeated tracking. Edelen \emph{et al.} trained neural-network surrogates on injector simulations and used them for multi-objective optimization with orders-of-magnitude speed-up~\cite{edelen2020}; Gupta \emph{et al.} built a measurement-based surrogate of the LCLS-II injector~\cite{gupta2021}; tree-based and neural models have been added to the operational toolsets of light sources~\cite{vera2019} and applied to muon cooling, laser-plasma sources and beam-dynamics forecasting~\cite{fol2022,brogren2025,rautela2024}. In most of these studies the model input is a set of machine settings. A second line of work predicts beam properties from upstream \emph{beam} measurements: virtual diagnostics infer the longitudinal phase space from non-invasive signals~\cite{emma2018,hanuka2021,bettoni2024,lundquist2026} with calibrated uncertainty~\cite{convery2021}, while tomographic and differentiable-simulation methods reconstruct transverse or four-dimensional phase space from measured projections~\cite{wolski2022,roussel2023,kim2024,kaiser2024}.

For a transport line that delivers a strongly focused beam through dispersive sections, two questions remain open: how accurately the IP beam can be computed quickly when the full distribution is known, and which beam information a learned model needs when it is not. This paper answers both for the electron driver of the MeV $\gamma\gamma$ collider and develops and validates the prediction components for the first two stages above. Specifically, we
\begin{enumerate}
\item construct a design-stage predictor that applies second-order transfer maps to every particle, and quantify its accuracy and speed against Elegant tracking;
\item construct a predictor for use in a digital twin that maps the means and rms widths of the entrance beam coordinates to the IP parameters, show that these beam moments are a better input than the injector settings, and test it with measurable inputs and input noise;
\item explain, parameter by parameter, which transport physics each approach captures, using linear $\sigma$-matrix transport as a reference.
\end{enumerate}
Connecting the twin to the real machine, including diagnostic calibration and experimental validation, is left for future work. We do not claim a new learning algorithm, and we do not compare with accuracies reported on other data sets.

% ======================================================== SYSTEM AND DATA
\section{Electron driver, transport line and simulations}\label{sec:system}
\subsection{Collider layout and studied segment}
\begin{figure*}[!t]
\centering\includegraphics[width=\textwidth]{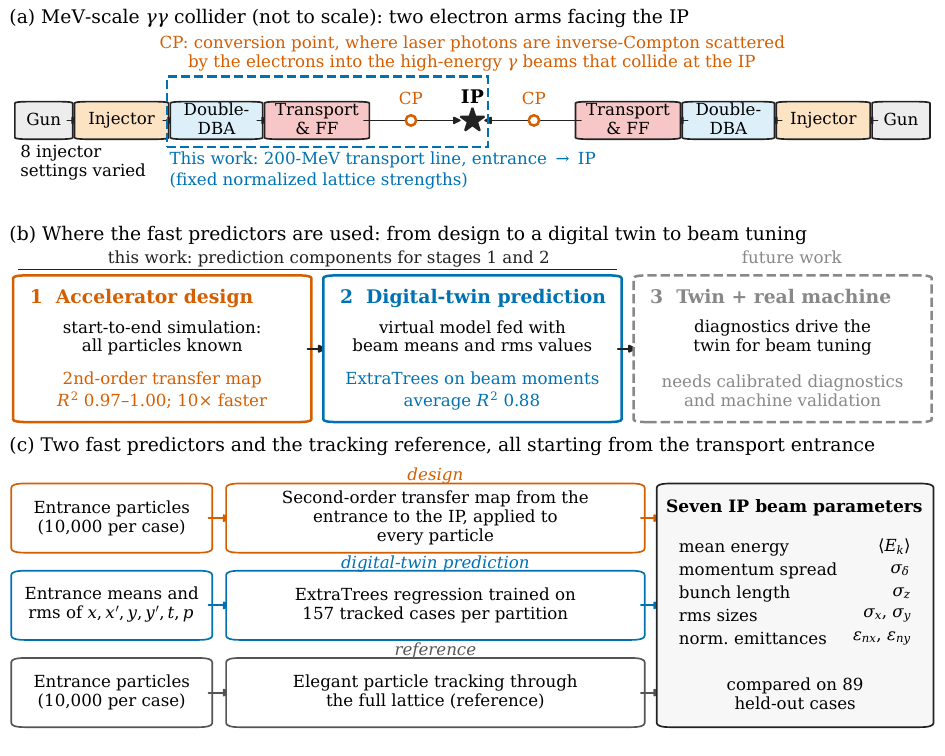}
\caption{Overview of the study. (a) Layout of the MeV $\gamma\gamma$ collider (not to scale): in each arm a gun and an injector produce a 200-MeV electron bunch, which a double double-bend achromat (Double-DBA) and a transport and final-focus (FF) channel deliver to the interaction point (IP). Laser photons are inverse-Compton scattered at the conversion points (CP, open circles) into the colliding $\gamma$ beams. The dashed box marks the studied transport line; the eight varied injector settings are RF phases, fields and solenoid strengths. (b) The three stages from accelerator design to a digital twin and to beam tuning on the real machine; this work develops and validates the prediction components for stages 1 and 2. (c) The two fast predictors and the tracking reference. Both predictors and the reference start from the same rematched entrance distribution and return the same seven IP parameters.}
\label{fig:overview}
\end{figure*}

Figure~\ref{fig:overview}(a) shows the collider layout. In each of the two arms, a photocathode gun and an injector produce a nominal 1-nC, 200-MeV electron bunch~\cite{zhang2026epjst}; a double-DBA section and a transport and final-focus channel then deliver it to the IP, where the two converted photon beams collide. The study covers one arm from the rematched transport entrance ($s=0$) to the electron tracking endpoint at the IP position ($s=32.63$~m). Photon conversion and photon collisions are not simulated. Figure~\ref{fig:overview}(b) places the two predictors in the development stages described above, and Fig.~\ref{fig:overview}(c) summarizes how each predictor works. Table~\ref{tab:design} lists the design context, the fixed conditions of this study and the observed range of each IP parameter.

\begin{table}[t]
\caption{Design context, fixed conditions and observed ranges. IP ranges are over the 89 evaluation cases; transmission is over all 200 cases.}
\label{tab:design}
\centering\footnotesize
\begin{tabular}{@{}p{0.34\linewidth}p{0.56\linewidth}@{}}\toprule
Item & Value \\\midrule
\multicolumn{2}{@{}l}{\emph{Design context}~\cite{zhang2026epjst}}\\
Beam energy, charge & 200~MeV, 1~nC\\
IP spot size & micrometre level\\
Vacuum chamber inner diameter & 6~mm (design)\\
\multicolumn{2}{@{}l}{\emph{Fixed in this study}}\\
Entrance Twiss & $\beta_x=3.013$~m, $\alpha_x=2.689$; $\beta_y=3.878$~m, $\alpha_y=0.427$\\
Downstream lattice & normalized strengths fixed; CSR, ISR, wakes off\\
Macroparticles & 10{,}000 per case\\
\multicolumn{2}{@{}l}{\emph{Observed at the IP}}\\
$\langle E_k\rangle$ & 195.9--204.2~MeV\\
$\sigma_\delta$ & 0.043--0.284\%\\
$\sigma_x$, $\sigma_y$ & 1.6--12.5~$\mu$m, 1.8--7.1~$\mu$m\\
$\sigma_z$ & 0.013--0.397~mm\\
$\epsilon_{nx}$, $\epsilon_{ny}$ & 0.97--24.6, 1.12--10.6~mm\,mrad\\
Transmission $T$ & 0.478--0.9999 (median 0.996)\\\bottomrule
\end{tabular}
\end{table}

\subsection{Transport line}\label{sec:line}
\begin{figure*}[!t]
\centering\includegraphics[width=\textwidth]{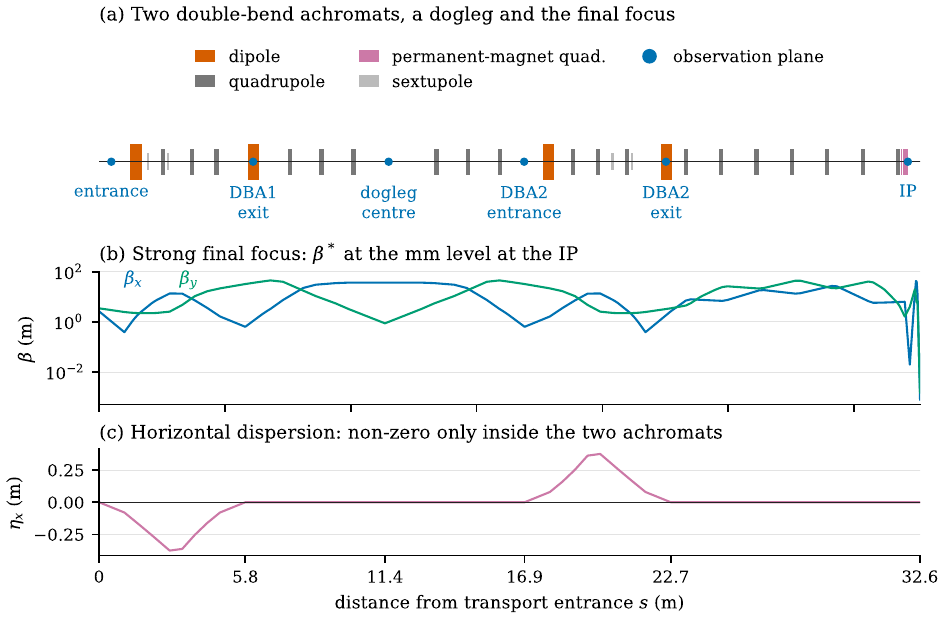}
\caption{Fixed transport line from the rematched entrance ($s=0$) to the IP. (a) Element layout and observation planes. (b) First-order beta functions. (c) Horizontal dispersion.}
\label{fig:beamline}
\end{figure*}

The line (Fig.~\ref{fig:beamline}) consists of two double-bend achromats (DBA1, DBA2) separated by a symmetric dogleg, followed by a ten-quadrupole final-focus channel whose last three magnets are permanent-magnet quadrupoles. Four sector bends of 0.158768~rad and four sextupoles are present. In the first-order reference optics the IP beta functions are $\beta_x^*=0.79$~mm and $\beta_y^*=1.50$~mm. The design vacuum chamber has an inner diameter of 6~mm.

The simulated lattice includes a historical particle-selection procedure that is not part of the confirmed design. The reference tracking and the map predictor use consistent selection conditions; details are given in the Supplementary Information.

\subsection{Start-to-end simulations}
Each case starts with ASTRA tracking of the gun and injector to 17~m with three-dimensional space charge and mirror charge~\cite{floettmann2017}, using 10{,}000 macroparticles. The distribution is converted to SDDS with all ordered rows retained and rematched to a common entrance Twiss condition (Table~\ref{tab:design}), so all cases enter the downstream line with the same nominal optics but different beam distributions. Elegant~\cite{borland2000,borland2023manual} then tracks the fixed lattice to the IP with \texttt{default\_order}=3, 50 slices and fourth-order integration per bend, and CSR, ISR and wakefields switched off. The quadrupole strengths are held fixed as normalized $K_1$ values with each case's own reference momentum. CSR calculations for four representative cases did not qualify as labels over this operating domain (Supplementary Information), so the downstream labels are radiation-free by construction rather than because CSR is negligible.

\begin{figure*}[!p]
\centering\includegraphics[width=\textwidth]{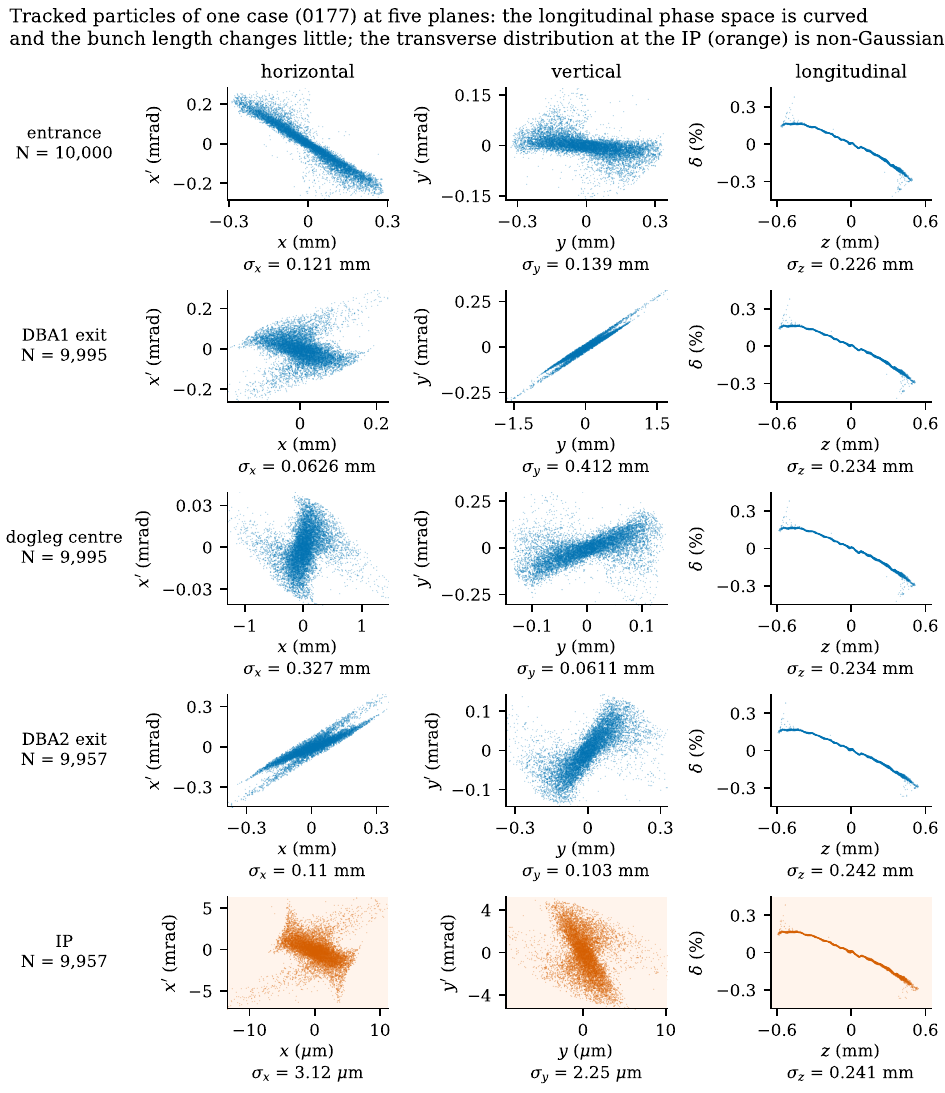}
\caption{Tracked macroparticles of one representative case (\texttt{0177}) at the entrance, the DBA1 exit, the dogleg centre, the DBA2 exit and the IP (shaded row). Columns show the horizontal, vertical and longitudinal phase spaces; $N$ is the number of surviving particles at each plane, and the rms size $\sigma_x$, $\sigma_y$ or bunch length $\sigma_z$ is given below each panel. The longitudinal coordinate is $z=\beta c(\langle t\rangle-t)$, with the bunch head at $z>0$. Transverse limits use the 0.5--99.5 percentile range; longitudinal limits are shared.}
\label{fig:phasespace}
\end{figure*}

Eight injector settings vary between the cases: three RF phases, three accelerating-field amplitudes and two solenoid fields (observed ranges in Supplementary Table~S4). The sampling law of the case generator was not retained, so we report the realized ranges. The retained records do not establish whether individual injector cases were separately optimized. The entrance longitudinal phase space is curved (Fig.~\ref{fig:phasespace}), and the entrance rms momentum spread ranges from 0.034 to 0.53\%. Figure~\ref{fig:phasespace} also shows that the transverse distribution at the IP is strongly non-Gaussian after the final focus.

\subsection{IP parameters and data partition}
Seven parameters are computed from the particles that reach the IP (Table~\ref{tab:targets}). The 200 cases were split five times into 157 training, 21 validation and 22 test cases (partition seeds 1, 2, 3, 42 and 123). The test sets overlap; their union contains 89 distinct cases, each of which is held out for every model whose test set includes it. Unless otherwise specified, the accuracy comparisons below use these 89 cases and the same Elegant results. Each partition places six of the eight cases with transmission below 0.80 in training, one in validation and one in test. A separately tracked set of 500 additional cases is used only as a consistency check, because its control-sampling history is incompletely documented.

\begin{table*}[!b]
\caption{Seven IP parameters computed from the surviving particles; $p=\beta\gamma$ is the momentum in units of $m_ec$.}
\label{tab:targets}
\centering\small
\begin{tabular}{@{}lll@{}}\toprule
Parameter & Unit & Definition \\\midrule
Mean kinetic energy $\langle E_k\rangle$ & MeV & $\left\langle(\sqrt{1+p^2}-1)m_ec^2\right\rangle$ \\
Relative momentum spread $\sigma_\delta$ & \% & $100\,\mathrm{std}[p/\langle p\rangle-1]$ \\
Horizontal, vertical rms size $\sigma_x,\sigma_y$ & $\mu$m & $10^6\,\mathrm{std}(x)$, $10^6\,\mathrm{std}(y)$ \\
Rms bunch length $\sigma_z$ & mm & $10^3c\,\mathrm{std}(t)$ \\
Normalized emittance $\epsilon_{nx},\epsilon_{ny}$ & mm\,mrad & $10^6\langle p\rangle\sqrt{\det\Sigma_{(x,x')}}$, likewise for $y$ \\\bottomrule
\end{tabular}
\end{table*}

% ================================================================= METHODS
\section{Fast predictors and evaluation}\label{sec:methods}
\subsection{Design stage: transfer maps}
Elegant~2026.2.0~\cite{elegant2026} provides the second-order transfer maps from the entrance to the IP and intermediate checkpoints,
\begin{equation}
 q'_i=C_i+\sum_jR_{ij}q_j+\sum_{j\geq k}T_{ijk}q_jq_k,
 \label{eq:map}
\end{equation}
with $q=(x,x',y,y',s,\delta)$, $s=\beta(p)ct$ and $\delta=p/p_{\rm ref}-1$; the stored $T_{ijk}$ multiply unique monomials. The predictor applies the maps to each of the 10{,}000 entrance particles: the same particle-selection conditions as in the reference tracking are applied at the corresponding locations, and the seven parameters are computed from the particles reaching the IP. Keeping only $R$ gives the first-order map. For this lattice the map coefficients are identical for all 200 cases, so in a design study with fixed normalized magnet strengths the maps need to be generated only once. The map evaluation was checked against an independent polynomial composition of the final-focus elements and against direct calls of the production code with exact test polynomials; third-order maps are not used because an independent quadrupole test did not reproduce their coefficients within tolerance (Supplementary Information).

\subsection{Digital-twin stage: ExtraTrees on entrance beam moments}
The learned predictor uses the mean and population rms width of each of the six entrance coordinates $(x,x',y,y',t,p)$, i.e.\ 12 beam moments that are closer to what diagnostics provide than a particle distribution. An extremely-randomized-trees (ExtraTrees) regressor~\cite{geurts2006} maps them to the seven IP parameters, with 64 trees, all inputs available at each split, a minimum of two samples per split and one per leaf, no depth limit and no bootstrap sampling. Inputs and targets are standardized with training-partition statistics. Five estimator seeds are trained in each of the five partitions (25 models); for each of the 89 cases the prediction is the mean over all models for which the case was held out. No hyperparameter was changed after test results were inspected. For comparison, the same learner was trained on the eight injector settings instead of the beam moments, and on the six means and 21 covariances of the entrance distribution.

\subsection{Reference transport models}
Linear $\sigma$-matrix transport propagates the six means $\mu$ and 21 covariances $\Sigma$ of the entrance distribution with $\mu'=C+R\mu$ and $\Sigma'=R\Sigma R^{\rm T}$, without particle losses. It shows what linear optics alone predicts from the beam moments. A second-order moment propagation under a Gaussian assumption performs worse and is reported in the Supplementary Information.

\subsection{Evaluation}
Unless otherwise specified, all predictors are scored on the same 89 cases against the same Elegant results. For each parameter we report the coefficient of determination $R^2=1-\sum_i(y_i-\hat y_i)^2/\sum_i(y_i-\bar y)^2$, the mean absolute error (MAE), the RMSE and the bias; $R^2=1$ is exact agreement and $R^2=0$ equals predicting the mean. The average $R^2$ is the unweighted mean over the seven parameters. Paired differences of absolute errors are given with 95\% intervals from 10{,}000 bootstrap replicates that resample cases. Transmission groups ($T<0.95$: 6 cases; $T\geq0.95$: 83 cases) are descriptive and were not used for fitting or selection.

\subsection{Use of AI tools}
OpenAI ChatGPT (historical version unavailable) and OpenAI Codex (recorded models \texttt{gpt-5.6-sol}, \texttt{gpt-5.6-terra}, \texttt{gpt-6-astra} and \texttt{gpt-6-sol}) assisted with restructuring, language revision, analysis code, figures and consistency checks. The physical assumptions, protocols and interpretation are the authors' responsibility; these tools are not authors.

% ================================================================= RESULTS
\section{Results}\label{sec:results}
\subsection{Entrance beam moments are a better input than injector settings}
\begin{figure*}[!t]
\centering\includegraphics[width=\textwidth]{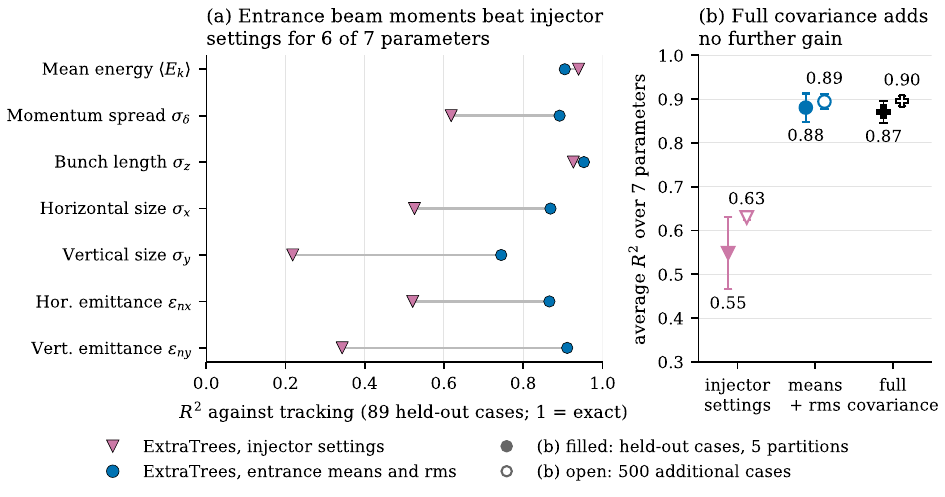}
\caption{Choice of input for the learned predictor. (a) $R^2$ of the same ExtraTrees learner trained on the eight injector settings or on the means and rms widths of the six entrance coordinates, evaluated on the same 89 held-out cases; each case prediction averages the models for which it was held out. (b) Average $R^2$ over the seven parameters under the original protocol (metrics averaged over the five partitions; bars are one partition standard deviation) for the held-out cases and the 500 additional cases, including the full entrance covariance.}
\label{fig:input}
\end{figure*}

The digital-twin predictor depends on which upstream information it receives. With the same ExtraTrees learner and the same held-out cases, replacing the eight injector settings by the 12 entrance beam moments raises the average $R^2$ from 0.585 to 0.877 (Fig.~\ref{fig:input}a; Supplementary Table~S30). The gain is largest for the transverse parameters: $\sigma_y$ rises from 0.22 to 0.74, $\epsilon_{ny}$ from 0.34 to 0.91 and $\epsilon_{nx}$ from 0.52 to 0.87. Only for the mean energy are the settings slightly better (0.940 versus 0.905). The beam moments describe the injector's actual response, so the model does not have to learn that response as well. The original partition-averaged protocol gives the same picture (0.548 versus 0.881 on held-out cases and 0.630 versus 0.894 on the 500 additional cases), and adding the full entrance covariance gives no further gain (0.871 and 0.895; Fig.~\ref{fig:input}b). A digital twin should therefore be supplied with measured beam moments rather than only with settings.

\subsection{Accuracy of the two fast predictors}
\begin{figure*}[!t]
\centering\includegraphics[width=\textwidth]{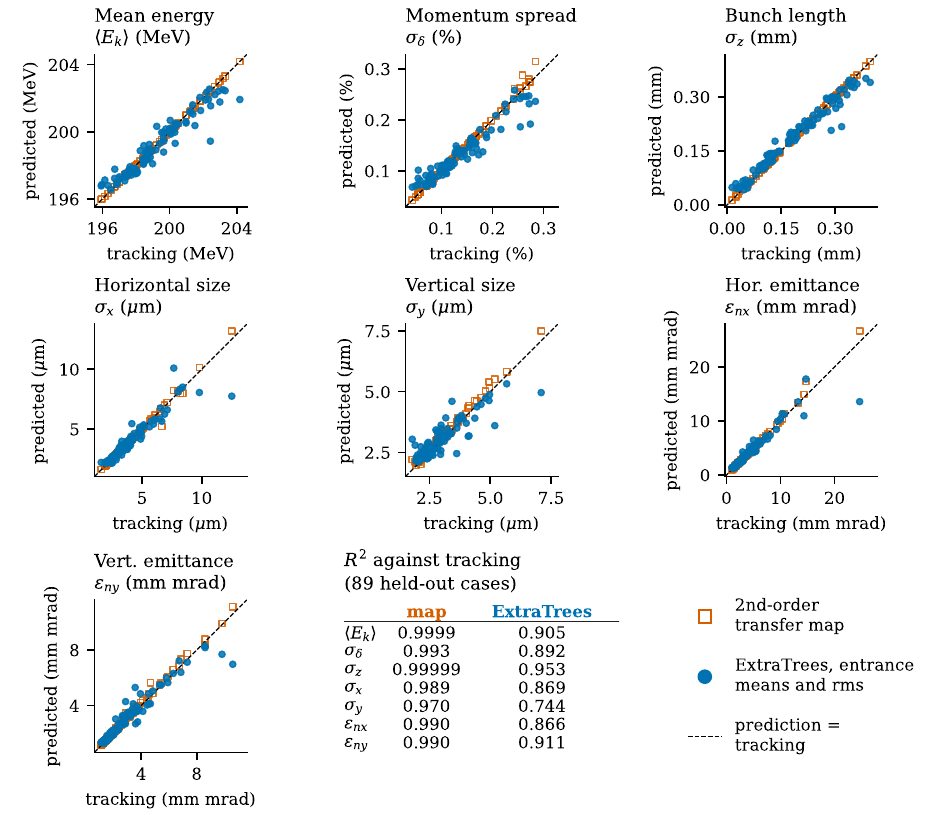}
\caption{Predicted versus tracked IP parameters for the 89 held-out cases. Open squares: second-order transfer map; filled circles: ExtraTrees on the entrance means and rms widths. The dashed line is prediction = tracking. The final panel lists $R^2$ of both predictors for each parameter.}
\label{fig:pred}
\end{figure*}

Figure~\ref{fig:pred} compares both predictors with Elegant tracking. The second-order map reproduces the tracked values for every parameter ($R^2=0.970$--1.000; average 0.990). ExtraTrees follows the trends with parameter-dependent scatter, from $R^2=0.744$ for $\sigma_y$ to 0.953 for $\sigma_z$ (average 0.877); its largest deviations in $\sigma_x$ and $\epsilon_{nx}$ occur in the few cases with the strongest particle loss (Supplementary Information). The paired bootstrap intervals of the absolute-error difference between the two predictors exclude zero for every parameter (Supplementary Table~S29); for $\epsilon_{nx}$, for example, the mean difference is 0.39~mm\,mrad (95\% interval 0.23--0.63).

\subsection{Which physics each predictor captures}
\begin{figure*}[!t]
\centering\includegraphics[width=\textwidth]{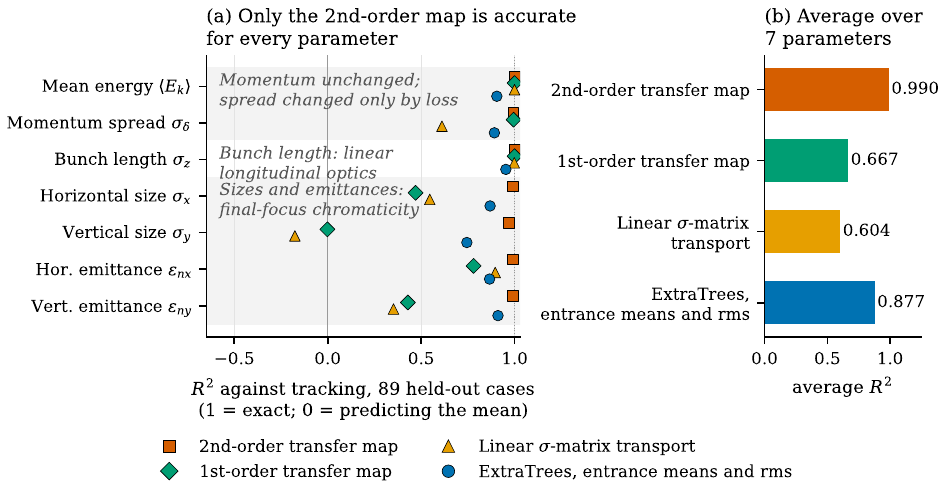}
\caption{Accuracy by parameter and model on the 89 held-out cases. (a) $R^2$ against Elegant tracking; shaded bands group the parameters by the physics that determines them. (b) Average $R^2$ over the seven parameters. Full tables with MAE, RMSE and bias in physical units are in Supplementary Table~S27.}
\label{fig:method}
\end{figure*}

Figure~\ref{fig:method} resolves the accuracy by parameter for the two predictors and the two transport references. Three groups appear.

\emph{Mean energy and momentum spread.} The line contains no accelerating structure, so individual particle momenta are conserved; changes in the ensemble momentum statistics arise from the simulated particle selection. Linear $\sigma$-matrix transport, which keeps every particle, therefore gives $R^2=0.9993$ for $\langle E_k\rangle$ but only 0.611 for $\sigma_\delta$, whereas the particle maps, which apply the selection, reach 0.993--0.994. ExtraTrees reaches 0.905 and 0.892.

\emph{Bunch length.} $\sigma_z$ follows from linear longitudinal transport acting on the time--momentum correlation; all transport routes reach $R^2\geq0.998$ and ExtraTrees 0.953. The 12 beam moments do not contain the time--momentum correlation, whereas the covariances used by $\sigma$-matrix transport do.

\emph{Transverse sizes and emittances.} Here linear optics is insufficient. Linear $\sigma$-matrix transport and the first-order map give $R^2=-0.18$ and 0.00 for $\sigma_y$ and are weak for the other transverse parameters. The second-order map, which includes the chromatic aberrations of the strong final focus and acts on the actual non-Gaussian distribution, reaches $R^2=0.970$--0.990 for all four. ExtraTrees, trained on tracked cases, learns much of this nonlinear response from the 12 beam moments ($R^2=0.744$--0.911) and is more accurate than both linear references for $\sigma_x$, $\sigma_y$ and $\epsilon_{ny}$; for $\epsilon_{nx}$ linear $\sigma$-matrix transport is slightly better (0.896 versus 0.866).

Neural networks fed with observations at five planes along the line or with phase-space images did not give consistent gains over a single set of beam moments (Supplementary Information).

\begin{figure*}[!t]
\centering\includegraphics[width=\textwidth]{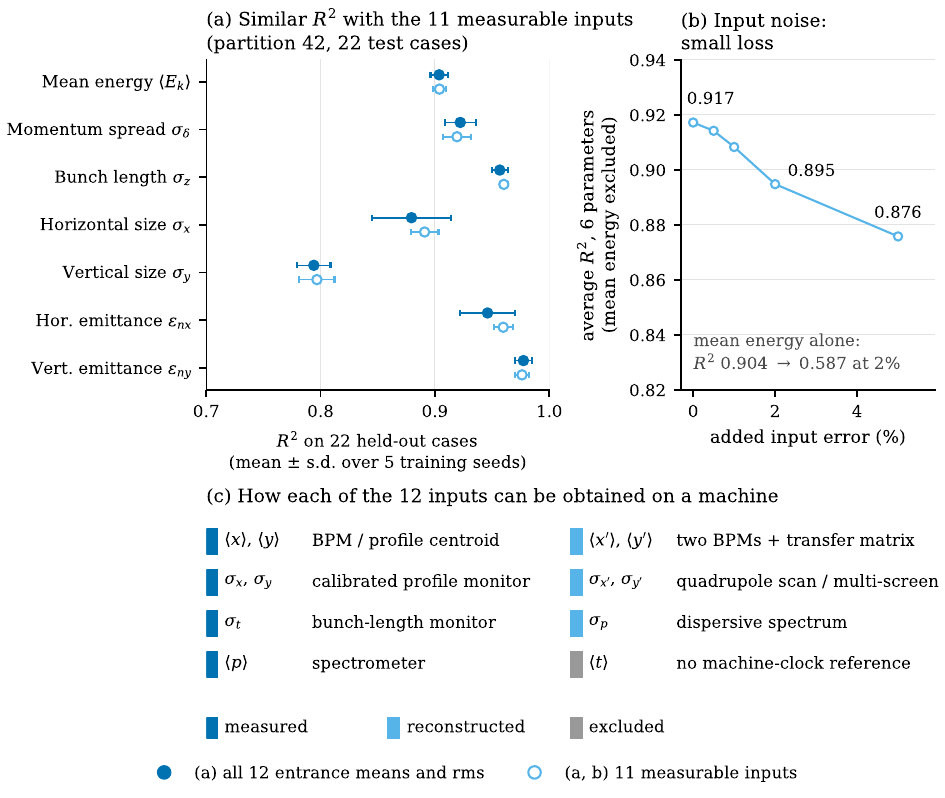}
\caption{Measurable inputs for the digital-twin predictor. (a) $R^2$ per parameter with all 12 entrance means and rms widths and with the 11 inputs that have an identified measurement or reconstruction path, on the 22 test cases of partition 42 (mean $\pm$ standard deviation over five training seeds). (b) Average $R^2$ of the six parameters other than the mean energy when random errors are added to the 11 inputs; positive quantities receive relative errors and signed quantities additive errors scaled from the training set. (c) How each input can be obtained on a machine. Details in Supplementary Tables~S21--S24.}
\label{fig:vd}
\end{figure*}

\begin{figure*}[!t]
\centering\includegraphics[width=\textwidth]{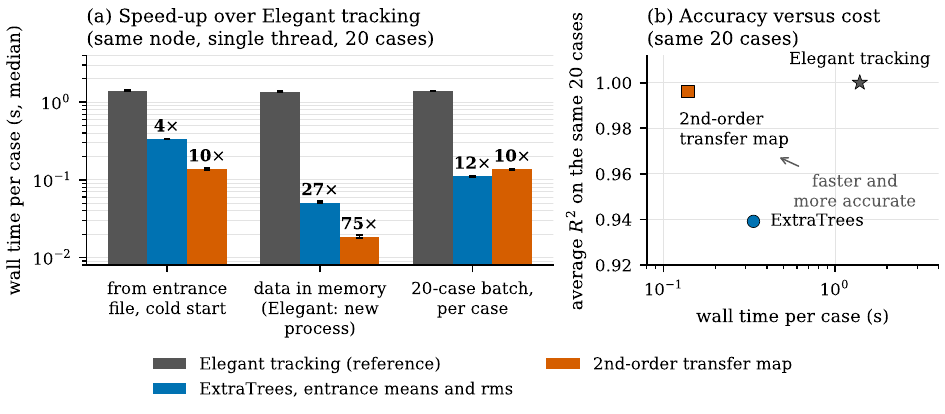}
\caption{Execution time on one compute node, single thread, for 20 cases of the additional set. (a) Median wall time per case for three usage modes; bars give the interquartile range and labels the paired speed-up relative to Elegant tracking. In the in-memory mode Elegant still starts a new tracking process from staged files. (b) Average $R^2$ on the same 20 cases versus wall time from the entrance file.}
\label{fig:timing}
\end{figure*}

\subsection{Measurable inputs for the digital twin}
A digital twin connected to a machine receives measured rather than simulated beam moments. We classified each of the 12 inputs by how it can be obtained (Fig.~\ref{fig:vd}c): centroids, rms sizes, bunch length and mean momentum are measured directly with beam-position monitors, profile monitors, a bunch-length monitor and a spectrometer; mean angles, angular spreads and momentum spread require standard reconstruction with transfer matrices, quadrupole scans or dispersive spectra~\cite{wiedemann2015,wolski2022}. Only the stored mean arrival time is excluded, because its simulation time origin has no defined mapping to a machine clock; this does not imply that the bunch arrival time cannot be measured. A model trained with the remaining 11 inputs, with otherwise unchanged settings, performs similarly to the full model. On the 22 test cases of partition 42 the average $R^2$ is 0.915 with 11 inputs and 0.911 with 12, and every per-parameter difference is smaller than the seed-to-seed standard deviation of the full-input model (Fig.~\ref{fig:vd}a). On the 500 additional cases the averages are 0.906 and 0.908; the largest per-parameter change there is for $\sigma_y$ (0.813 versus 0.821). With random input errors added at prediction time, the average $R^2$ of the six parameters other than the mean energy decreases from 0.917 to 0.895 at 2\% and to 0.876 at 5\% (Fig.~\ref{fig:vd}b). The mean energy is more sensitive: its $R^2$ falls from 0.904 to 0.587 at 2\%, so that the seven-parameter average falls from 0.915 to 0.851. An accurately calibrated momentum scale is therefore required for the energy prediction. The 2\% level refers to the perturbation rule of the Supplementary Information (relative errors for positive quantities, additive errors for signed quantities), not to a common accuracy of all diagnostics. These results use simulated inputs and do not replace validation with machine diagnostics.

\subsection{Execution time}
The predictors and Elegant were timed on one compute node with one thread, on 20 cases of the additional 500-case set, with one excluded warm-up, ten repeats and rotated order (Fig.~\ref{fig:timing}). Starting from the entrance particle file and returning the seven parameters, the median times per case are 1.390~s for Elegant, 0.334~s for ExtraTrees (25 models loaded) and 0.138~s for the second-order map including map generation, i.e.\ 4 and 10 times faster than tracking. With data and models already in memory the speed-ups are 27 and 75; in this mode Elegant still starts a new process from staged files. In a 20-case batch, ExtraTrees loads its models once and is 12 times faster, whereas the map, regenerated for every case in this measurement, is 10 times faster; because the maps are identical for all cases of this configuration, a design study can reuse them. On these 20 cases the average $R^2$ is 0.939 for ExtraTrees and 0.996 for the map. When the particle distribution is available, the map is therefore both faster and more accurate than the learned model; the learned model is needed when only beam moments are known. Setup times and all boundaries are listed in Supplementary Table~S32.

% ================================================================= DISCUSSION
\section{Discussion}\label{sec:discussion}
The two predictors provide the prediction components for the first two stages of the collider's beam-modelling chain. In the design stage, the full simulated distribution is available, and the second-order map reproduces Elegant for all seven IP parameters at a tenth of the cost, without training data. It can therefore replace downstream tracking when many injector states are screened, provided the maps are generated for the lattice actually used. For digital-twin use, only beam moments are available. ExtraTrees on the entrance means and rms widths then gives an average $R^2$ of 0.88, which is far better than a model that uses only the injector settings, performs similarly with the 11 measurable inputs and, except for the mean energy, is robust to moderate input noise.

The comparison also shows what a moment-based description misses. Mean energy, momentum spread and bunch length are largely determined by momentum conservation and linear longitudinal optics, while the transverse IP parameters reflect the nonlinear transport of the simulated particle distribution. A combination in which transport physics supplies the energy and longitudinal parameters and the learned model the transverse ones is suggested by Fig.~\ref{fig:method} but was also not tested.

The third stage, connecting the twin to the real accelerator for beam tuning, requires steps beyond this work: calibrated diagnostics for the 11 inputs; validation of the predictors against measured IP parameters; and models that include fixed physical magnet gradients, CSR, ISR and wakefields, which are excluded from the present labels. A frozen model loses accuracy when the final focus is changed after its last input (Supplementary Information), so each lattice configuration requires its own maps and, for the learned model, revalidation or retraining. Further limits are that all inputs and labels are simulated, and that the 500-case set is a consistency check rather than a blind test. Finally, the historical particle selection in the reference simulations does not represent the confirmed design. The predictors therefore require validation for the actual line configuration; the learned model may require retraining if the changed conditions affect its accuracy.

% ================================================================= CONCLUSION
\section{Conclusion}\label{sec:conclusion}
We developed two fast predictors of the electron-beam parameters at the IP of a MeV $\gamma\gamma$ collider for its 200-MeV transport line. For accelerator design, second-order transfer maps applied to every particle reproduce Elegant tracking for all seven parameters ($R^2=0.97$--1.00) and are ten times faster. For use in a digital twin, an ExtraTrees model on the entrance means and rms widths reaches an average $R^2$ of 0.88, far above a model using injector settings, and performs similarly with eleven measurable inputs; under a 2\% input perturbation the six parameters other than the mean energy change little, whereas the mean energy requires an accurately calibrated momentum scale. Together they provide the IP-beam evaluation for the collider design and the prediction layer of its digital twin; before they are used for the design, both should be validated on simulations representing the confirmed line configuration.

\FloatBarrier
\backmatter
\bmhead{Acknowledgements}
This work is supported by the Shanghai Action Plan for Science, Technology and Innovation (Grant No.~24JD1402800), the Guangdong Provincial Key Laboratory of Advanced Particle Detection Technology (Grant No.~2024B1212010005), the Guangdong Provincial Key Laboratory of Gamma-Gamma Collider and Its Comprehensive Applications (Grant No.~2024KSYS001), and the National Key Research and Development Program of China (Grant No.~2023YFA1607200).
The authors also acknowledge the computational support provided by the
High-performance Computing Public Platform (Shenzhen Campus) of Sun Yat-sen
University.

\bmhead{Statements and Declarations}
\bmhead{Funding}
The funding sources and grant numbers are listed in the Acknowledgements.

\bmhead{Competing interests}
The authors declare no competing interests relevant to this article.

\bmhead{Data availability}
The processed data, raw ASTRA/SDDS particle files, transfer-map coefficients and trained models are
available from the corresponding author upon reasonable request.

\bmhead{Code availability}
The analysis and plotting scripts are available from the corresponding author upon
reasonable request.

\bmhead{Supplementary information}
Online Resource~1 contains the control ranges, tracking conventions, map checks, complete parameter tables and paired intervals, transmission groups, the second-order Gaussian moment propagation, the neural-network representation comparisons, the measurable-input and optics-perturbation studies, the CSR qualification study and the execution-time boundaries.
\bibliography{references}
\end{document}

\begin{center}
{\large\bfseries Supplementary Information (Online Resource 1)\par}
\medskip
{\Large\bfseries Fast prediction of electron-beam parameters using transfer maps and machine learning in a digital model of a MeV $\gamma\gamma$ collider\par}
\medskip
{\itshape The European Physical Journal C\par}
\medskip
Jiawei Zhang$^{1}$, Yongsheng Huang$^{1}$, Yuancun Nie$^{1}$, Qiang Gu$^{2}$, Yuan Chen$^{3}$, Yuanjie Bi$^{1,*}$\par
\smallskip
{\small $^{1}$School of Science, Shenzhen Campus of Sun Yat-sen University, Shenzhen 518107, China\par
$^{2}$Shanghai Advanced Research Institute, Chinese Academy of Sciences, Shanghai, China\par
$^{3}$Wuhan University, Wuhan, China\par}
\smallskip
{\small $^{*}$Corresponding author: Yuanjie Bi, \href{mailto:biyj@mail.sysu.edu.cn}{biyj@mail.sysu.edu.cn}\par}
\medskip
{\small\begin{tabular}{ll}
Jiawei Zhang & \href{mailto:zhangjw267@mail2.sysu.edu.cn}{zhangjw267@mail2.sysu.edu.cn}\\
Yongsheng Huang & \href{mailto:huangysh59@mail.sysu.edu.cn}{huangysh59@mail.sysu.edu.cn}\\
Yuancun Nie & \href{mailto:nieyc5@mail.sysu.edu.cn}{nieyc5@mail.sysu.edu.cn}\\
Qiang Gu & \href{mailto:guq@sari.ac.cn}{guq@sari.ac.cn}\\
Yuan Chen & \href{mailto:chenyuan79@whu.edu.cn}{chenyuan79@whu.edu.cn}\\
Yuanjie Bi & \href{mailto:biyj@mail.sysu.edu.cn}{biyj@mail.sysu.edu.cn}
\end{tabular}\par}
\end{center}
\raggedbottom
\renewcommand{\thetable}{S\arabic{table}}
\renewcommand{\theHtable}{S\arabic{table}}
\setcounter{table}{0}
\renewcommand{\thefigure}{S\arabic{figure}}
\renewcommand{\theHfigure}{S\arabic{figure}}
\setcounter{figure}{0}

This Supplementary Information gives the representation definitions, full
target-level results, uncertainty conventions, optical-map checks, and
supporting diagnostic and transport sensitivity studies. The associated CSV
tables retain per-partition and per-seed metrics and are available from the
corresponding author upon reasonable request.

Short labels are used in the tables of this Supplementary Information.
Table~\ref{tab:notation} gives their correspondence with the names used in
the main text.

\begin{table}[htbp]
\caption{Correspondence between Supplementary labels and main-text names}
\label{tab:notation}
\centering\small
\begin{tabular}{@{}ll@{}}\toprule
Supplementary label & Main-text name \\\midrule
Particle-R2 & second-order transfer map with particle selection \\
Particle-R1 & first-order transfer map with particle selection \\
Moment-R1 & linear $\sigma$-matrix transport \\
Moment-R2 (G) & second-order moment propagation, Gaussian assumption \\
ET-12, Entrance-12 & ExtraTrees on the entrance means and rms widths \\
Complete-28 & ExtraTrees on the entrance means and covariances \\
Settings & ExtraTrees on the eight injector settings \\
Entrance-VD & ExtraTrees on the 11 measurable inputs \\
M1, Distributed-VD & neural model on five observation planes (full / measurable) \\
RCOL & zero-length particle cut, $|x|\le2$~mm, in the simulated lattice (not part of the design) \\
$W_{\rm IP}$ & electron tracking endpoint at the IP \\\bottomrule
\end{tabular}
\end{table}

\section{IP ensemble and beam-property definitions}
\label{sec:s-system}

\subsection{Controls, compact entrance state, and targets}

The IP analysis uses 200 start-to-end cases. Three radio-frequency (RF) phases,
three accelerating-field amplitudes, and two solenoid-field amplitudes vary in
the retained injector ensemble. The model boundary is the rematched transport
entrance to the stored $W_{\rm IP}$/IP endpoint. The 12-variable entrance vector
contains the mean and population standard deviation of $x$, $x'$, $y$, $y'$, $t$,
and $p$. The neural scalar branch adds particle
count,
rms relative momentum spread $\sigma_\delta$, and the dimensionless quantity
$e_{\rm Twiss,max}$, the maximum absolute residual in the numerical consistency
check performed after the common entrance Twiss condition is applied. The seven
IP targets are common to every route: mean kinetic energy, relative momentum
spread, horizontal and vertical rms size, bunch length, and normalized
horizontal and vertical emittance. A relative momentum spread is reported as a
percentage-point quantity; it is not called an energy spread without the
corresponding conversion. No formal acceptance threshold is introduced when a
source does not provide one.

The entrance input fields and their ranges are listed in
Table~\ref{tab:s2-scalars}.

\begin{table}[htbp]
\caption{Fifteen entrance variables used by the neural scalar branch. Ranges are
observed over the 200-case IP ensemble}
\label{tab:s2-scalars}
\centering\scriptsize
\renewcommand{\arraystretch}{0.95}
\begin{tabular}{@{}rllrrp{0.42\textwidth}@{}}\toprule
No. & Field & Unit & Min. & Max. & Definition \\ \midrule
1 & \texttt{entrance\_n\_particles} & count & 10000 & 10000 & Transport-entrance
SDDS row count; constant simulation quantity. \\
2 & \texttt{entrance\_x\_mean\_m} & m & $-3.2795\!\times\!10^{-6}$ &
$1.9189\!\times\!10^{-6}$ & Mean $x$; centroid diagnostic analogue. \\
3 & \texttt{entrance\_x\_std\_m} & m & $8.0694\!\times\!10^{-5}$ &
$3.8593\!\times\!10^{-4}$ & Rms $x$; profile diagnostic analogue. \\
4 & \texttt{entrance\_xp\_mean} & rad & $-8.1281\!\times\!10^{-7}$ &
$5.1134\!\times\!10^{-6}$ & Mean $x'$; requires optics-based reconstruction. \\
5 & \texttt{entrance\_xp\_std} & rad & $7.6646\!\times\!10^{-5}$ &
$3.6459\!\times\!10^{-4}$ & Rms $x'$; periodic multi-profile analogue. \\
6 & \texttt{entrance\_y\_mean\_m} & m & $-2.1817\!\times\!10^{-6}$ &
$4.9842\!\times\!10^{-6}$ & Mean $y$; centroid diagnostic analogue. \\
7 & \texttt{entrance\_y\_std\_m} & m & $9.1187\!\times\!10^{-5}$ &
$4.3822\!\times\!10^{-4}$ & Rms $y$; profile diagnostic analogue. \\
8 & \texttt{entrance\_yp\_mean} & rad & $-1.7411\!\times\!10^{-6}$ &
$4.0430\!\times\!10^{-6}$ & Mean $y'$; requires optics-based reconstruction. \\
9 & \texttt{entrance\_yp\_std} & rad & $2.5556\!\times\!10^{-5}$ &
$1.2107\!\times\!10^{-4}$ & Rms $y'$; periodic multi-profile analogue. \\
10 & \texttt{entrance\_t\_mean\_s} & s & $5.6757\!\times\!10^{-8}$ &
$5.6778\!\times\!10^{-8}$ & Mean arrival-time coordinate; reference dependent.
\\
11 & \texttt{entrance\_t\_std\_s} & s & $1.5750\!\times\!10^{-13}$ &
$1.2748\!\times\!10^{-12}$ & Rms duration; specialized diagnostic analogue. \\
12 & \texttt{entrance\_p\_mean} & 1 & 384.3853 & 400.5800 & Mean normalized
momentum $p/(m_e c)$. \\
13 & \texttt{entrance\_p\_std} & 1 & 0.13145 & 2.06469 & Rms normalized
momentum. \\
14 & \texttt{entrance\_delta\_std} & 1 & $3.3925\!\times\!10^{-4}$ &
$5.3060\!\times\!10^{-3}$ & Rms of $p/\langle p\rangle-1$. \\
15 & \texttt{max\_after\_twiss\_abs\_error} & 1 & $5.7732\!\times\!10^{-15}$ &
$3.4625\!\times\!10^{-11}$ & Numerical residual from the entrance-optics
interface. \\
\bottomrule
\end{tabular}
\end{table}
The common target definitions are given in Table~\ref{tab:s2-targets}.

\begin{table}[htbp]
\caption{IP target schema. All targets are computed from \texttt{W\_IP.out}; the
IP is excluded from model inputs and input normalization}
\label{tab:s2-targets}
\centering\scriptsize
\begin{tabular}{@{}lllp{0.57\textwidth}@{}}\toprule
Target & Symbol & Unit & Definition \\ \midrule
\texttt{Ekin\_mean\_MeV} & $\langle E_{\rm kin}\rangle$ & MeV &
$\langle(\sqrt{1+p^2}-1)m_e c^2\rangle$. \\
\texttt{sigma\_delta\_percent} & $\sigma_\delta$ & \% &
$100\,\mathrm{std}[p/\langle p\rangle-1]$. \\
\texttt{sigmax\_mm} & $\sigma_{x,\rm IP}$ & mm & $1000\,\mathrm{std}(x)$. \\
\texttt{sigmay\_mm} & $\sigma_{y,\rm IP}$ & mm & $1000\,\mathrm{std}(y)$. \\
\texttt{sigmaz\_mm} & $\sigma_{z,\rm IP}$ & mm & $1000c\,\mathrm{std}(t)$. \\
\texttt{emitnx\_mm\_mrad} & $\epsilon_{nx,\rm IP}$ & mm\,mrad & $10^6\langle
p\rangle\sqrt{\langle x^2\rangle\langle x'^2\rangle-\langle xx'\rangle^2}$ after
centering. \\
\texttt{emitny\_mm\_mrad} & $\epsilon_{ny,\rm IP}$ & mm\,mrad & $10^6\langle
p\rangle\sqrt{\langle y^2\rangle\langle y'^2\rangle-\langle yy'\rangle^2}$ after
centering. \\
\bottomrule
\end{tabular}
\end{table}
The realized injector-control ranges are reported in
Table~\ref{tab:s2-controls}.
For the primary 200-case cohort, the case generator, intended sampling law,
prescribed control bounds, nominal/reference control vector, and random seed
are unavailable. Table~\ref{tab:s2-controls} therefore reports realized extrema,
not prescribed sampling bounds. The particle-source law used for the CSR
qualification in Sec.~\ref{sec:s-csr} is known separately.

\begin{table}[htbp]
\caption{Observed ranges of the eight controls in the 200-case IP ensemble}
\label{tab:s2-controls}
\centering\scriptsize
\begin{tabular}{@{}llrr@{}}\toprule
Control & Unit & Observed min & Observed max \\ \midrule
$\phi(1)$ & deg & -0.972780 & 3.979955 \\
$\phi(2)$ & deg & 214.122602 & 223.980768 \\
$\phi(3)$ & deg & -83.908591 & -68.027274 \\
$\mathrm{maxe}(1)$ & MV/m & 105.053713 & 114.997729 \\
$\mathrm{maxe}(2)$ & MV/m & 18.530983 & 22.496783 \\
$\mathrm{maxe}(3)$ & MV/m & 15.047664 & 18.946122 \\
$\mathrm{maxb}(1)$ & T & 0.198168 & 0.231762 \\
$\mathrm{maxb}(2)$ & T & 0.036006 & 0.051961 \\
\bottomrule
\end{tabular}
\end{table}

\subsection{Tracking and particle conventions}

ASTRA (A Space Charge Tracking Algorithm) tracks the injector to 17~m with
three-dimensional space charge and
mirror charge. Injector statistics omit the absolute reference row in the ASTRA
particle
file, whereas the ASTRA-to-Self-Describing Data Sets (SDDS) conversion
retained all 10,000 ordered rows. Every downstream route and stored target uses
the same
converted files. As a sensitivity check, removing ordered row zero changes the
seven IP
quantities by no more than $2.10\times10^{-4}$ relatively.

At the transport interface, the stored analysis distributions use the common
entrance
Twiss condition, $\beta_x=3.01288$~m, $\alpha_x=2.68854$,
$\beta_y=3.87758$~m, and $\alpha_y=0.426671$. The same entrance optics are used
for all downstream cases.

The numerical sensitivity to particle and moment conventions is summarized in
Table~\ref{tab:definition-sensitivity}.

\begin{table}[t]
\caption{Maximum absolute relative changes under alternative beam-property
definitions and ordered-row removal. Row-removal entries refer to the IP}
\label{tab:definition-sensitivity}
\centering
\small
\begin{tabular}{@{}lc@{}}
\toprule
Comparison or IP quantity & Maximum relative change \\
\midrule
$c\sigma_t$ versus particle-$\beta$ $\sigma_z$ (entrance) & $3.0\times10^{-6}$
\\
$c\sigma_t$ versus particle-$\beta$ $\sigma_z$ (IP) & $3.0\times10^{-6}$ \\
$\epsilon_{nx}$: trace space versus momentum covariance (entrance) &
$1.0491\times10^{-2}$ \\
$\epsilon_{ny}$: trace space versus momentum covariance (entrance) &
$4.837\times10^{-3}$ \\
$\epsilon_{nx}$: trace space versus momentum covariance (IP) &
$1.454\times10^{-3}$ \\
$\epsilon_{ny}$: trace space versus momentum covariance (IP) &
$1.173\times10^{-3}$ \\
Ordered-row removal: $\langle E_k\rangle$ & $6.02\times10^{-7}$ \\
Ordered-row removal: $\sigma_\delta$ & $9.52\times10^{-5}$ \\
Ordered-row removal: $\sigma_x$ & $1.04\times10^{-4}$ \\
Ordered-row removal: $\sigma_y$ & $1.05\times10^{-4}$ \\
Ordered-row removal: $\sigma_z$ & $7.46\times10^{-5}$ \\
Ordered-row removal: $\epsilon_{nx}$ & $2.01\times10^{-4}$ \\
Ordered-row removal: $\epsilon_{ny}$ & $2.10\times10^{-4}$ \\
\bottomrule
\end{tabular}
\end{table}

The simulation \textsc{elegant} driver uses one run-control step and
\texttt{default\_order=3}. Each bend uses 50 slices and fourth-order
integration. Coherent synchrotron radiation (CSR) and incoherent synchrotron
radiation (ISR) switches are zero, and the lattice contains no wakefield
element. Four sextupoles provide nonlinear correction. These settings define
the downstream magnetic transport used for the representation comparisons.
The lattice holds normalized magnetic strengths fixed, including quadrupole
$K_1$, with each case's reference momentum used consistently for particles and
maps. Fixed physical-gradient hardware at varying reference momentum would
require a different strength convention.

The stored observation planes and absolute longitudinal coordinates are listed
in Table~\ref{tab:watchpoint-map}.

\begin{table}[t]
\caption{Longitudinal locations in the retained \textsc{elegant} tracking files.
The transport entrance defines $s=0$. The final-focus entrance and
second-achromat exit are coincident markers}
\label{tab:watchpoint-map}
\centering
\small
\begin{tabular}{@{}rllrc@{}}
\toprule
Index & Name & Beamline boundary & $s$ (m) & Model input \\
\midrule
0 & entrance & transport entrance / first-achromat entrance & 0.000000 & yes \\
1 & First-achromat exit & after the second bend of the first achromat & 5.808427
& yes \\
2 & dogleg center & symmetry point of the dogleg & 11.358957 & yes \\
3 & Second-achromat entrance & before the first bend of the second achromat &
16.909487 & yes \\
4 & Second-achromat exit & after the second bend of the second achromat &
22.717914 & yes \\
5 & final-focus entrance & duplicate of index 4 & 22.717914 & excluded \\
6 & IP & end of final-focus transport & 32.625214 & excluded \\
\bottomrule
\end{tabular}
\end{table}

The second-achromat-exit and final-focus-entrance files contain identical
particle arrays
for all 200 cases after SDDS conversion. The five-state routes use the five
unique non-IP
states at indices 0--4; index 6 is the IP prediction plane.

Particle survival at the observation planes is summarized in
Table~\ref{tab:survival-summary}.

\begin{table}[t]
\caption{Particle survival relative to the 10,000-particle transport entrance.
First-achromat exit, dogleg center, and second-achromat entrance share the same
counts; second-achromat exit, the stored final-focus entrance, and the IP also
share the same counts}
\label{tab:survival-summary}
\centering
\small
\begin{tabular}{@{}lcccc@{}}
\toprule
Watchpoint & Mean & Minimum & Median & Maximum \\
\midrule
Entrance & 1.00000 & 1.0000 & 1.00000 & 1.0000 \\
First-achromat exit & 0.98540 & 0.5708 & 0.99910 & 1.0000 \\
Dogleg center & 0.98540 & 0.5708 & 0.99910 & 1.0000 \\
Second-achromat entrance & 0.98540 & 0.5708 & 0.99910 & 1.0000 \\
Second-achromat exit & 0.97539 & 0.4780 & 0.99635 & 0.9999 \\
IP & 0.97539 & 0.4780 & 0.99635 & 0.9999 \\
\bottomrule
\end{tabular}
\end{table}

Eight cases have IP survival below 0.80. Each partition places six of those
cases in training, one in validation, and one in testing, within total counts
of 157/21/22. Test subsets overlap and their union contains 89 unique cases.
Partition standard deviations are therefore descriptive sensitivity measures.

\FloatBarrier

\section{Representation definitions and entrance results}
\subsection{Map-coordinate moment input and Gaussian closure}
The Moment-R1 and Moment-R2 routes use $q=(x,x',y,y',s,\delta)$, where
$s=\beta(p)ct$ and $\delta=p/p_{\rm ref}-1$. The propagated state consists
of six means and the 21 unique entries of the population covariance matrix.
The particle count is constant at 10,000 and is not used by either propagation
function. These 27 varying map-coordinate moments are therefore not identical
to the legacy Complete-28 machine-learning representation described below:
that representation stores the particle count, means of $(x,x',y,y',t,p)$,
and covariances of $(x,x',y,y',\zeta,\delta)$, with
$\zeta=\beta_0c(t-\langle t\rangle)$ and $\delta=p/\langle p\rangle-1$.
Transforming between the two requires the nonlinear particle-dependent
velocity factor and a reference-momentum change, so finite moments alone do
not provide an exact general conversion. We consequently call the physics
baseline ``map-coordinate moments'' and do not claim it is the legacy
Complete-28 input.

For each output coordinate, write the quadratic map as
$f_i(q)=C_i+R_iq+q^TH_iq$, with symmetric $H_i$ formed from the stored unique
lower-triangular monomials. A diagonal coefficient multiplies $q_a^2$;
an off-diagonal stored coefficient multiplies $q_aq_b$ once and is split
equally between $H_{i,ab}$ and $H_{i,ba}$. Under the Gaussian closure
$q\sim\mathcal{N}(\mu,\Sigma)$,
\begin{align}
\mathbb{E}[f_i]&=C_i+R_i\mu+\mu^TH_i\mu+\operatorname{tr}(H_i\Sigma),\\
\operatorname{Cov}(f_i,f_j)&=b_i^T\Sigma b_j+
2\operatorname{tr}(H_i\Sigma H_j\Sigma),\qquad
b_i=R_i+2H_i\mu.
\end{align}
This is exact for the moments of a quadratic function of a Gaussian input,
but is a closure approximation for the non-Gaussian beam and does not include
particle-level selection. The historical formula check used 200,000 correlated Gaussian samples and produced maximum mean and covariance residuals of 0.00149 and 0.00334 in synthetic units. A separate deterministic check now calls the production propagation function, testing all 28 individual monomials and a mixed random polynomial against exact Gaussian quadrature. Its largest mean/covariance residual is $1.42\times10^{-14}$. This algebraic test has no Monte Carlo sampling error; it verifies the formula and coefficient convention, rather than the Gaussian approximation for real beams.

The map-coordinate longitudinal spread is reported as $\sigma_s$, whereas
the common IP target is $c\,\mathrm{std}(t)$. Since $s=\beta(p)ct$, these are
not algebraically identical. The retained convention-sensitivity check found
a maximum relative difference of $3.0\times10^{-6}$ in the entrance/IP
cohort; this small difference is documented as an approximation, not an exact
unit conversion. The remaining moment observables use mean normalized
momentum $p_{\rm ref}(1+\mu_\delta)$, the propagated momentum variance, and
the transverse covariance determinants as implemented in the archived
prediction function.

\subsection{Complete 28-variable entrance state}

The complete entrance vector contains the particle count, the six means of
$(x,x',y,y',t,p)$, and the upper triangle of the population covariance matrix
for $(x,x',y,y',\zeta,\delta)$. The longitudinal coordinate within the bunch is
$\zeta=\beta_0c[t-\langle t\rangle]$ and
$\delta=p/\langle p\rangle-1$, where $\beta_0$ is the entrance
reference-particle velocity divided by $c$. The 21 covariance entries include
$\operatorname{cov}(x,\delta)$, $\operatorname{cov}(x',\delta)$, all
cross-plane correlations, and the longitudinal correlations. The entrance
count is constant at 10,000, so the vector has 28 stored positions and 27
varying quantities. All cases share the same transport-entrance Twiss condition.
Consequently, the within-plane transverse second-moment blocks vary within a
common optics boundary, while Complete-28 adds cross-plane, momentum-coupled,
and longitudinal correlations.

The compact tabular-baseline configuration is specified in main-text Sec.~3.2. The historical Complete-28 comparison retains its original protocol; its partition-averaged score is not reinterpreted as a common-case ensemble score. Five
estimator seeds are evaluated for ExtraTrees within each partition; Ridge is
deterministic.

The complete entrance-state results are resolved by target in
Table~\ref{tab:s-complete-entrance}.

\begin{table}[t]
\caption{Target-level results for the complete 28-variable entrance
representation. The particle count is constant at 10,000 in this ensemble.
Here nMAE denotes normalized mean absolute error. ExtraTrees values average
five estimator seeds within each of the five partitions; ridge has one
deterministic fit per partition}
\label{tab:s-complete-entrance}
\centering
\small
\begin{tabular}{@{}lcccc@{}}
\toprule
& \multicolumn{2}{c}{Ridge} & \multicolumn{2}{c}{ExtraTrees} \\
\cmidrule(lr){2-3}\cmidrule(l){4-5}
Target & $R^2$ & nMAE & $R^2$ & nMAE \\
\midrule
$\langle E_k\rangle$ & $0.99913\pm0.00018$ & $0.02019\pm0.00339$ &
$0.87989\pm0.07064$ & $0.24632\pm0.08119$ \\
$\sigma_\delta$ & $0.39267\pm0.41491$ & $0.48731\pm0.06047$ &
$0.88585\pm0.04082$ & $0.24394\pm0.07667$ \\
$\sigma_x$ & $0.51154\pm0.47171$ & $0.31239\pm0.09339$ & $0.85938\pm0.07675$ &
$0.21448\pm0.04810$ \\
$\sigma_y$ & $0.40681\pm0.39438$ & $0.41907\pm0.07611$ & $0.72152\pm0.10666$ &
$0.31280\pm0.02208$ \\
$\sigma_z$ & $0.97315\pm0.01418$ & $0.12560\pm0.02229$ & $0.94077\pm0.03852$ &
$0.17510\pm0.02100$ \\
$\epsilon_{nx}$ & $0.76507\pm0.22162$ & $0.21610\pm0.08305$ &
$0.89076\pm0.06646$ & $0.17761\pm0.06757$ \\
$\epsilon_{ny}$ & $0.82575\pm0.13114$ & $0.21071\pm0.04611$ &
$0.92079\pm0.04721$ & $0.15137\pm0.03472$ \\
\midrule
Macro mean & $0.69630\pm0.20341$ & $0.25591\pm0.04077$ & $0.87128\pm0.02514$ &
$0.21737\pm0.01712$ \\
\bottomrule
\end{tabular}
\end{table}

For the entrance baselines, input and target z-scores use training-partition
statistics, with a denominator replaced by one when below $10^{-12}$. The
primary and external Entrance-12/Complete-28 evaluations use the same input
standardization. The matched neural comparisons instead replace only exactly
zero standard deviations. Full settings and entrance partition metrics are
available as CSV tables from the corresponding author upon reasonable request.

\subsection{Physical descriptors and the common entrance context}
\begin{table}[H]
\caption{Seventeen low-order beam descriptors computed at each of the five
unique non-IP states. Population moments use divisor $N$; $\delta=p/\langle
p\rangle-1$. All six SDDS source columns must be finite before any feature is
computed}
\label{tab:r4-moments}
\centering
\small
\begin{tabular}{rllll}
\toprule
No. & Quantity & Formula & Unit & SDDS column(s) \\
\midrule
1 & n\_particles & $N$ & count & $x,x',y,y',t,p$ \\
2 & mean\_x & $\langle x\rangle$ & m & $x$ \\
3 & std\_x & $\sqrt{\langle(x-\langle x\rangle)^2\rangle}$ & m & $x$ \\
4 & mean\_xp & $\langle x'\rangle$ & rad & $x'$ \\
5 & std\_xp & $\sqrt{\langle(x'-\langle x'\rangle)^2\rangle}$ & rad & $x'$ \\
6 & cov\_x\_xp & $\langle(x-\langle x\rangle)(x'-\langle x'\rangle)\rangle$ &
m\,rad & $x,x'$ \\
7 & mean\_y & $\langle y\rangle$ & m & $y$ \\
8 & std\_y & $\sqrt{\langle(y-\langle y\rangle)^2\rangle}$ & m & $y$ \\
9 & mean\_yp & $\langle y'\rangle$ & rad & $y'$ \\
10 & std\_yp & $\sqrt{\langle(y'-\langle y'\rangle)^2\rangle}$ & rad & $y'$ \\
11 & cov\_y\_yp & $\langle(y-\langle y\rangle)(y'-\langle y'\rangle)\rangle$ &
m\,rad & $y,y'$ \\
12 & mean\_t & $\langle t\rangle$ & s & $t$ \\
13 & std\_t & $\sqrt{\langle(t-\langle t\rangle)^2\rangle}$ & s & $t$ \\
14 & mean\_p & $\langle p\rangle$ & 1 & $p$ \\
15 & std\_p & $\sqrt{\langle(p-\langle p\rangle)^2\rangle}$ & 1 & $p$ \\
16 & std\_delta & $\sqrt{\langle(\delta-\langle\delta\rangle)^2\rangle}$ & 1 &
$p$ \\
17 & cov\_t\_delta & $\langle(t-\langle
t\rangle)(\delta-\langle\delta\rangle)\rangle$ & s & $t,p$ \\
\bottomrule
\end{tabular}
\end{table}
The terminal descriptor input has shape $[B,1,17]$, the five-state input
$[B,5,17]$, the common entrance branch $[B,15]$, and the output $[B,7]$, where
$B$ is batch size. The distributed five-state model is denoted M1 in the
supporting studies below. It uses the first five stored observation indices; the
terminal model selects index 4 only. Neither includes beamline position or
position increments. Masked pooling averages valid encoded states. With one
valid state it reduces to the terminal encoding using the same learned
parameters.

The common entrance branch includes the simulation count and numerical Twiss
residual defined above, identically in both arms. The comparison therefore
measures the contribution of earlier observations given the same entrance
context, shared encoder, and equal-weight pooling.

\subsection{Images and physical metadata}
Terminal image input has shape $[B,3,64,64]$. For each projection, both axes use
case-local 0.5th and 99.5th percentiles. Equal percentile bounds use the image
constructor's deterministic fallback. Histogram bin counts are converted to
float32, transformed by log1p, and divided by the channel maximum. No
total-count normalization, density normalization, or extra five-percent axis
padding is applied. Points outside the two-dimensional bounds are omitted.

\begin{table}[H]
\caption{Ordered terminal metadata schema, identical for the image-plus-metadata
and metadata-only routes. Bounds are lower/upper pairs}
\label{tab:metadata}
\centering\small
\begin{tabular}{lll}\toprule
Indices & Quantities & Units\\\midrule
1--2 & $x$ lower/upper percentile bounds & m\\
3--4 & $x'$ lower/upper bounds & dimensionless slope\\
5--6 & $y$ lower/upper bounds & m\\
7--8 & $y'$ lower/upper bounds & dimensionless slope\\
9--10 & $t$ lower/upper bounds & s\\
11--12 & $\delta$ lower/upper bounds & dimensionless\\
13 & $\langle p\rangle$, defining $\delta$ & $m_ec$ normalization\\
14 & terminal macroparticle count & count\\
15--17 & maximum log counts in $(x,x')$, $(y,y')$, $(t,\delta)$ &
dimensionless\\
\bottomrule\end{tabular}
\end{table}

The image construction with metadata reproduces the 200 stored terminal image
tensors exactly. The external 500-case images are constructed on demand from
particle files; the two constructors give identical arrays for all 700 cases.
Metadata contains bounds, normalization scales, and coarse occupancy
information, while omitting tails outside the bounds, full covariances, and IP
quantities. The terminal comparison therefore tests the contribution of pixels
given both scale and occupancy information. It contains no terminal image-only
arm.

\section{Matched networks, preprocessing, and evaluation}
\begin{table}[H]
\caption{Neural routes used in the matched observation and terminal
representation comparisons. Every row retains the same entrance context and
seven-target objective}
\label{tab:networks}
\centering\small
\begin{tabular}{llll}\toprule
Route & Observation encoding & Aggregation/fusion & Parameters\\\midrule
Five-state & shared $17\to407\to64$ & masked mean; entrance fusion & 74,949\\
Terminal A & same descriptor encoder & one valid state; same fusion & 74,949\\
Terminal B & CNN64 + metadata $17\to64\to64$ & $128\to64$; entrance fusion &
88,519\\
Terminal C & metadata $17\to64\to64$ & entrance fusion & 46,823\\
\bottomrule\end{tabular}
\end{table}

The entrance encoder is $15\to32\to32$ with SiLU and dropout 0.1. The fused
output head is $96\to384\to7$. The metadata encoder uses linear/SiLU/linear
operations. B uses the shared image encoder with 16 then 32 channels, GroupNorm,
SiLU, residual blocks, global average and maximum pooling, and a 64-dimensional
embedding. The five-state comparator and A are frozen for the B/C experiment. B and C each
contribute 25 fitted models, with the parameter counts shown in
Table~\ref{tab:networks}.

The partition seeds are 1, 2, 3, 42, and 123; training seeds are 1--5. All
routes use AdamW with learning rate $10^{-3}$, weight decay $10^{-4}$, batch
size 16, at most 100 epochs, validation early-stopping patience 15,
learning-rate reduction factor 0.5 with patience 5, and gradient-norm clipping
at 1. The objective gives equal weight to the seven standardized targets.
Checkpoints are selected by validation MSE.

The neural moment and metadata standard deviations are computed in float64 using
training cases only; exactly zero values use denominator one. Entrance means and
target means/standard deviations are computed in float32, and entrance standard
deviations in float64. B/C reuse A's entrance and target normalization
statistics. Image statistics are per channel and fitted only from terminal
pixels of training cases, computed in float64 and applied with float32 image
arithmetic. Saved-model prediction reproducibility was checked for all 50 B/C
fits: the largest difference was $3.82\times10^{-6}$ in
target-standard-deviation units, below the predefined $2\times10^{-5}$
tolerance.

\subsection{Error definitions and bootstrap units}
For residual $r_i=\hat y_i-y_i$, MAE is $\langle|r|\rangle$, RMSE is
$\sqrt{\langle r^2\rangle}$, and bias is $\langle r\rangle$. Each target $R^2$
uses the variance of the evaluation cases, while nMAE uses the target population
standard deviation of the corresponding training partition.

For the main neural estimator, metrics are evaluated for each fitted model and
averaged over seeds and partitions. The test union comprises 89 cases; a case
may enter several partition metrics. Case bootstrap draws resampling weights
over the unique eligible cases and applies each weight to all appearances of
that case, retaining its seeds, partitions, and targets together. Each partition
metric is renormalized by its resampled weight total before averaging. There are
10,000 replicates. The intervals quantify case-sampling uncertainty conditional
on the fitted models.

The observation comparison uses 99.375\% intervals for four targets times two
cohorts. The terminal representation comparison uses 99.7917\% intervals for
three pairs times four targets times two cohorts; its six transverse-mean
comparisons use 99.1667\%. Ordinary 95\% intervals are also reported.
These intervals assess paired error differences; an equivalence test would
require an application-derived margin.

The unique-OOS nMAE sensitivity estimator first ensembles seed predictions
within each partition. For each OOS appearance, the absolute error of that
ensemble is divided by the corresponding partition's training-target population
standard deviation. These normalized absolute errors are averaged over the OOS
appearances of each physical case, and the resulting case errors are averaged
equally over the 89 unique cases. Thus prediction ensembling precedes error
calculation only within a partition; across partitions this statistic averages
normalized errors.

The supplementary physical-unit ensemble metrics use a separate aggregation:
each case's genuinely OOS partition predictions are averaged before evaluation
on the 89 unique cases. For the supporting external comparisons, the model-wise estimator computes
each model metric; the supplementary seed-ensemble estimator averages seed
predictions within each partition before evaluating its 500 cases. The
explicit-optics analysis uses the latter external estimator and the primary
unique-OOS prediction ensemble.

\section{Complete matched-observation and representation results}
The matched-observation tables report MAE, RMSE, bias, $R^2$, nMAE, all seven
targets, individual fits, and paired intervals. Their principal transverse mean
is $-0.00661$ in the primary cohort and $-0.00622$ externally when a positive
value would favor five observations. The unique-OOS counterpart is $-0.00549$
with 95\% interval $[-0.02909,0.02587]$. None of the four transverse targets
supports a positive history increment in both cohorts after multiplicity
adjustment. Several RMSE point estimates favor the five-state route; the absence
of a consistent MAE gain applies to the matched shared-encoder, equal-weight
pooled representation.

\begin{table}[H]
\caption{Matched observation comparison on primary OOS. Per-model metrics
averaged over seeds and partitions; original target units}
\label{tab:history-primary200}
\centering\small
\begin{tabular}{lllllll}
\toprule
Target & Route & MAE & RMSE & Bias & $R^2$ & nMAE \\\midrule
$\langle E_k\rangle$ & Five-state & 0.091742 & 0.13364 & -0.014885 & 0.9933 &
0.047747 \\
$\langle E_k\rangle$ & Terminal & 0.12778 & 0.20647 & -0.040608 & 0.98587 &
0.067027 \\
$\sigma_\delta$ & Five-state & 0.003128 & 0.0052288 & -0.000699 & 0.99151 &
0.054243 \\
$\sigma_\delta$ & Terminal & 0.0028007 & 0.0041805 & -0.000236 & 0.99377 &
0.048416 \\
$\sigma_x$ & Five-state & 0.000282 & 0.000383 & -4.79e-06 & 0.95013 & 0.1577 \\
$\sigma_x$ & Terminal & 0.000259 & 0.00037 & -2.13e-05 & 0.95462 & 0.1451 \\
$\sigma_y$ & Five-state & 0.000322 & 0.000498 & -1.71e-05 & 0.68041 & 0.32475 \\
$\sigma_y$ & Terminal & 0.000311 & 0.000548 & -2.1e-05 & 0.57411 & 0.31405 \\
$\sigma_z$ & Five-state & 0.0059445 & 0.0082392 & 0.000421 & 0.99155 & 0.060924
\\
$\sigma_z$ & Terminal & 0.0066467 & 0.0094457 & -7.72e-05 & 0.98869 & 0.068281
\\
$\epsilon_{nx}$ & Five-state & 0.31853 & 0.46283 & -0.023621 & 0.98016 &
0.097573 \\
$\epsilon_{nx}$ & Terminal & 0.31103 & 0.5294 & -0.077609 & 0.97476 & 0.095893
\\
$\epsilon_{ny}$ & Five-state & 0.35442 & 0.58419 & 0.016477 & 0.89466 & 0.16955
\\
$\epsilon_{ny}$ & Terminal & 0.34855 & 0.64266 & 0.011453 & 0.86277 & 0.16811 \\
\bottomrule\end{tabular}
\end{table}
\begin{table}[H]
\caption{Matched observation comparison on external 500. Per-model metrics
averaged over seeds and partitions; original target units}
\label{tab:history-external500}
\centering\small
\begin{tabular}{lllllll}
\toprule
Target & Route & MAE & RMSE & Bias & $R^2$ & nMAE \\\midrule
$\langle E_k\rangle$ & Five-state & 0.099187 & 0.14537 & -0.010941 & 0.99346 &
0.051525 \\
$\langle E_k\rangle$ & Terminal & 0.12638 & 0.19628 & -0.018267 & 0.98911 &
0.066283 \\
$\sigma_\delta$ & Five-state & 0.0027967 & 0.0049887 & -0.000174 & 0.99263 &
0.04841 \\
$\sigma_\delta$ & Terminal & 0.0026285 & 0.0041067 & 0.000225 & 0.99504 &
0.045332 \\
$\sigma_x$ & Five-state & 0.000278 & 0.000403 & -1.14e-05 & 0.94967 & 0.15596 \\
$\sigma_x$ & Terminal & 0.000262 & 0.000383 & -4.64e-06 & 0.95432 & 0.14733 \\
$\sigma_y$ & Five-state & 0.000305 & 0.000442 & -2e-05 & 0.83797 & 0.30867 \\
$\sigma_y$ & Terminal & 0.000287 & 0.000436 & 1.15e-07 & 0.84203 & 0.28966 \\
$\sigma_z$ & Five-state & 0.0060491 & 0.0083884 & 0.000319 & 0.99194 & 0.062042
\\
$\sigma_z$ & Terminal & 0.006441 & 0.0089394 & 0.000599 & 0.99093 & 0.066143 \\
$\epsilon_{nx}$ & Five-state & 0.36183 & 0.62124 & 0.024442 & 0.96504 & 0.11131
\\
$\epsilon_{nx}$ & Terminal & 0.36852 & 0.62604 & 0.036534 & 0.96417 & 0.11353 \\
$\epsilon_{ny}$ & Five-state & 0.34447 & 0.56853 & -0.011569 & 0.94143 & 0.16518
\\
$\epsilon_{ny}$ & Terminal & 0.34521 & 0.6101 & 0.0035749 & 0.93282 & 0.16572 \\
\bottomrule\end{tabular}
\end{table}
\FloatBarrier
The terminal tables retain all seven quantities. A is physical descriptors, B
pixels with metadata, and C metadata alone. Each target is reported in its
original physical units. The longitudinal quantities provide consistency checks.
\begin{table}[H]
\caption{Terminal representation metrics on the primary OOS cases. Metrics are
computed for each model before averaging seeds and partitions. Original units:
MeV, percentage points, mm, and mm mrad as defined in the main text}
\label{tab:terminal-primary200}
\centering\small
\begin{tabular}{lllllll}
\toprule
Target & Route & MAE & RMSE & Bias & $R^2$ & nMAE \\\midrule
$\langle E_k\rangle$ & A & 0.12778 & 0.20647 & -0.040608 & 0.98587 & 0.067027 \\
$\langle E_k\rangle$ & B & 0.37654 & 0.46492 & -0.0029425 & 0.93597 & 0.19752 \\
$\langle E_k\rangle$ & C & 0.19863 & 0.25506 & -0.0089732 & 0.97775 & 0.10394 \\
$\sigma_\delta$ & A & 0.0028007 & 0.0041805 & -0.000236 & 0.99377 & 0.048416 \\
$\sigma_\delta$ & B & 0.010438 & 0.0141 & -0.0022073 & 0.94361 & 0.18009 \\
$\sigma_\delta$ & C & 0.0090982 & 0.01209 & -0.000789 & 0.95479 & 0.15682 \\
$\sigma_x$ & A & 0.000259 & 0.00037 & -2.13e-05 & 0.95462 & 0.1451 \\
$\sigma_x$ & B & 0.000382 & 0.000616 & -4.43e-05 & 0.88225 & 0.21529 \\
$\sigma_x$ & C & 0.000341 & 0.000507 & 3.19e-06 & 0.91579 & 0.19111 \\
$\sigma_y$ & A & 0.000311 & 0.000548 & -2.1e-05 & 0.57411 & 0.31405 \\
$\sigma_y$ & B & 0.000334 & 0.0005 & 2.12e-05 & 0.69406 & 0.33834 \\
$\sigma_y$ & C & 0.000298 & 0.000413 & -3.69e-06 & 0.77934 & 0.30139 \\
$\sigma_z$ & A & 0.0066467 & 0.0094457 & -7.72e-05 & 0.98869 & 0.068281 \\
$\sigma_z$ & B & 0.016567 & 0.020787 & 0.0033394 & 0.95374 & 0.16991 \\
$\sigma_z$ & C & 0.015077 & 0.018913 & -0.000237 & 0.96117 & 0.15472 \\
$\epsilon_{nx}$ & A & 0.31103 & 0.5294 & -0.077609 & 0.97476 & 0.095893 \\
$\epsilon_{nx}$ & B & 0.51234 & 0.87189 & -0.054245 & 0.94074 & 0.1596 \\
$\epsilon_{nx}$ & C & 0.41957 & 0.67385 & -0.01512 & 0.96291 & 0.13028 \\
$\epsilon_{ny}$ & A & 0.34855 & 0.64266 & 0.011453 & 0.86277 & 0.16811 \\
$\epsilon_{ny}$ & B & 0.37066 & 0.59271 & 0.019678 & 0.88156 & 0.17763 \\
$\epsilon_{ny}$ & C & 0.35682 & 0.53797 & 0.025601 & 0.89117 & 0.17147 \\
\bottomrule\end{tabular}
\end{table}
\begin{table}[H]
\caption{Terminal representation metrics on the external 500 cases. Metrics are
computed for each model before averaging seeds and partitions. Original units:
MeV, percentage points, mm, and mm mrad as defined in the main text}
\label{tab:terminal-external500}
\centering\small
\begin{tabular}{lllllll}
\toprule
Target & Route & MAE & RMSE & Bias & $R^2$ & nMAE \\\midrule
$\langle E_k\rangle$ & A & 0.12638 & 0.19628 & -0.018267 & 0.98911 & 0.066283 \\
$\langle E_k\rangle$ & B & 0.402 & 0.50937 & -0.035777 & 0.93065 & 0.21058 \\
$\langle E_k\rangle$ & C & 0.20428 & 0.26885 & -0.014815 & 0.97891 & 0.10684 \\
$\sigma_\delta$ & A & 0.0026285 & 0.0041067 & 0.000225 & 0.99504 & 0.045332 \\
$\sigma_\delta$ & B & 0.010095 & 0.013804 & -0.000684 & 0.94624 & 0.17397 \\
$\sigma_\delta$ & C & 0.0083859 & 0.011617 & -0.000326 & 0.96202 & 0.14454 \\
$\sigma_x$ & A & 0.000262 & 0.000383 & -4.64e-06 & 0.95432 & 0.14733 \\
$\sigma_x$ & B & 0.000329 & 0.000479 & -4.77e-05 & 0.92877 & 0.18475 \\
$\sigma_x$ & C & 0.000302 & 0.000456 & -4.97e-06 & 0.93544 & 0.16958 \\
$\sigma_y$ & A & 0.000287 & 0.000436 & 1.15e-07 & 0.84203 & 0.28966 \\
$\sigma_y$ & B & 0.000315 & 0.000445 & 5.34e-06 & 0.83534 & 0.31851 \\
$\sigma_y$ & C & 0.000278 & 0.000388 & -3.71e-05 & 0.87471 & 0.28121 \\
$\sigma_z$ & A & 0.006441 & 0.0089394 & 0.000599 & 0.99093 & 0.066143 \\
$\sigma_z$ & B & 0.017387 & 0.022681 & 0.00071 & 0.94943 & 0.17829 \\
$\sigma_z$ & C & 0.015583 & 0.01998 & -6.88e-05 & 0.95892 & 0.15987 \\
$\epsilon_{nx}$ & A & 0.36852 & 0.62604 & 0.036534 & 0.96417 & 0.11353 \\
$\epsilon_{nx}$ & B & 0.45862 & 0.75342 & -0.023401 & 0.94815 & 0.14169 \\
$\epsilon_{nx}$ & C & 0.3914 & 0.64927 & 0.01886 & 0.96161 & 0.12099 \\
$\epsilon_{ny}$ & A & 0.34521 & 0.6101 & 0.0035749 & 0.93282 & 0.16572 \\
$\epsilon_{ny}$ & B & 0.38807 & 0.68775 & -0.02427 & 0.91455 & 0.18641 \\
$\epsilon_{ny}$ & C & 0.34689 & 0.57551 & -0.045102 & 0.94011 & 0.16653 \\
\bottomrule\end{tabular}
\end{table}
\FloatBarrier
\begin{table}[H]
\caption{Paired terminal-representation improvements on primary OOS. Positive
favors the first route. Adjusted coverage is 99.7917\% for targets and 99.1667\%
for the transverse mean}
\label{tab:paired-primary200}
\centering\small
\begin{tabular}{lllll}
\toprule
Pair & Target & $\Delta$ nMAE & 95\% interval & Adjusted interval \\\midrule
A--B & $\sigma_x$ & +0.07019 & [+0.02946, +0.11918] & [+0.01117, +0.15280] \\
A--B & $\sigma_y$ & +0.02429 & [-0.03877, +0.08475] & [-0.07730, +0.11893] \\
A--B & $\epsilon_{nx}$ & +0.06371 & [+0.03091, +0.10020] & [+0.01547, +0.12917]
\\
A--B & $\epsilon_{ny}$ & +0.00953 & [-0.03653, +0.04943] & [-0.06687, +0.06904]
\\
A--B & Transverse mean & +0.04193 & [+0.01967, +0.06410] & [+0.01320, +0.07287]
\\
B--C & $\sigma_x$ & -0.02418 & [-0.05311, +0.00021] & [-0.07262, +0.01032] \\
B--C & $\sigma_y$ & -0.03695 & [-0.09172, +0.01316] & [-0.12231, +0.03721] \\
B--C & $\epsilon_{nx}$ & -0.02932 & [-0.05565, -0.00792] & [-0.07404, +0.00147]
\\
B--C & $\epsilon_{ny}$ & -0.00617 & [-0.03300, +0.01793] & [-0.04775, +0.03183]
\\
B--C & Transverse mean & -0.02415 & [-0.05190, -0.00096] & [-0.06386, +0.00537]
\\
A--C & $\sigma_x$ & +0.04601 & [+0.01739, +0.07616] & [+0.00451, +0.09711] \\
A--C & $\sigma_y$ & -0.01265 & [-0.09768, +0.05606] & [-0.15797, +0.08926] \\
A--C & $\epsilon_{nx}$ & +0.03439 & [+0.00682, +0.06637] & [-0.00752, +0.08743]
\\
A--C & $\epsilon_{ny}$ & +0.00336 & [-0.04298, +0.04365] & [-0.07512, +0.06418]
\\
A--C & Transverse mean & +0.01778 & [-0.01268, +0.04477] & [-0.02366, +0.05247]
\\
\bottomrule\end{tabular}
\end{table}
\begin{table}[H]
\caption{Paired terminal-representation improvements on external 500. Positive
favors the first route. Adjusted coverage is 99.7917\% for targets and 99.1667\%
for the transverse mean}
\label{tab:paired-external500}
\centering\small
\begin{tabular}{lllll}
\toprule
Pair & Target & $\Delta$ nMAE & 95\% interval & Adjusted interval \\\midrule
A--B & $\sigma_x$ & +0.03742 & [+0.02847, +0.04664] & [+0.02322, +0.05253] \\
A--B & $\sigma_y$ & +0.02885 & [+0.01094, +0.04654] & [-0.00010, +0.05744] \\
A--B & $\epsilon_{nx}$ & +0.02816 & [+0.02031, +0.03618] & [+0.01632, +0.04093]
\\
A--B & $\epsilon_{ny}$ & +0.02069 & [+0.00572, +0.03622] & [-0.00137, +0.04794]
\\
A--B & Transverse mean & +0.02878 & [+0.02068, +0.03671] & [+0.01794, +0.03985]
\\
B--C & $\sigma_x$ & -0.01516 & [-0.02316, -0.00722] & [-0.02791, -0.00263] \\
B--C & $\sigma_y$ & -0.03730 & [-0.05335, -0.02151] & [-0.06299, -0.01208] \\
B--C & $\epsilon_{nx}$ & -0.02070 & [-0.02961, -0.01199] & [-0.03504, -0.00792]
\\
B--C & $\epsilon_{ny}$ & -0.01988 & [-0.03127, -0.00886] & [-0.03707, -0.00305]
\\
B--C & Transverse mean & -0.02326 & [-0.03084, -0.01555] & [-0.03377, -0.01309]
\\
A--C & $\sigma_x$ & +0.02225 & [+0.01320, +0.03115] & [+0.00830, +0.03648] \\
A--C & $\sigma_y$ & -0.00846 & [-0.02432, +0.00727] & [-0.03367, +0.01584] \\
A--C & $\epsilon_{nx}$ & +0.00746 & [-0.00201, +0.01704] & [-0.00755, +0.02187]
\\
A--C & $\epsilon_{ny}$ & +0.00082 & [-0.01186, +0.01293] & [-0.01832, +0.01950]
\\
A--C & Transverse mean & +0.00552 & [-0.00231, +0.01321] & [-0.00480, +0.01575]
\\
\bottomrule\end{tabular}
\end{table}
\FloatBarrier

\subsection{Aggregation sensitivity, tails, and transmission}
In the primary unique-OOS normalized-error sensitivity analysis, the A--B
transverse increment is 0.02332 with 95\% interval $[-0.00107,0.04808]$.
Horizontal size and horizontal emittance retain positive ordinary 95\% target
intervals, but their more conservative family intervals include zero. The
strength of the paired evidence therefore depends on aggregation. Seed-ensemble
and unique-case results are available in the interval tables from the
corresponding author upon reasonable request.

Primary vertical-size RMSE is approximately $0.548$, $0.500$, and $0.413~\mu$m
for A/B/C, and vertical-emittance RMSE is 0.643, 0.593, and 0.538~mm\,mrad.
These rankings differ from some MAE comparisons. No cases were removed because
of prediction errors, and no additional seed, hyperparameter, or architecture
selection followed the test results.

The transmission groups $T<0.8$, $0.8\leq T<0.95$, and $T\geq0.95$ remain
descriptive subsets. Small strata can have one case in a test partition; $R^2$
is then undefined. The CSV strata table preserves these missing values
and reports physical-unit errors.

\section{Terminal physical references and map consistency}
\subsection{Population and longitudinal definitions}
Stable particle identifiers match between the terminal and IP snapshots for all
700 cases. Sorting and matching use these identifiers.
Recomputing terminal input moments on the IP survivor set therefore leaves them
unchanged. The entrance particle files lack corresponding stable identifiers, so
the earlier entrance-to-IP population-selection contribution cannot be
quantified by the same method.

For dimensionless normalized total momentum $p$, the particle-exact mean energy
is $m_ec^2\langle\sqrt{1+p^2}-1\rangle$. The finite-moment direct approximation
is $m_ec^2(\sqrt{1+\langle p\rangle^2}-1)$. Its mean
absolute discrepancy is approximately $1.37\times10^{-9}$~MeV in the primary
cohort and $1.42\times10^{-9}$~MeV externally. Exact particle momenta are
conserved in the terminal segment. Direct bunch-length transport gives
$R^2=0.999874$ and 0.999869; the corresponding first-order values are 0.999880
and 0.999875. These extremely small errors must also be distinguished from
storage quantization when physics is compared with float32 neural labels.

The optics analysis uses particle-exact IP observables for every arm. The
terminal-representation experiment instead uses the same stored HDF5 truth for
all compact routes and its R2 reference. Its full-state table retains both the
exact-particle and HDF5 residuals. Near-zero energy or momentum-spread reference
errors are not used as divisors in performance ratios. The reported normalized
emittance is mean-$p$ times the geometric trace-space emittance, using the
definition in the main text.

\subsection{Linear and second-order segment maps}
The terminal segment is the 9.9073-m final-focus lattice with quadrupoles,
drifts, zero-kick correctors, monitors, and markers. The cumulative matrices
give
\[
 R_{\rm seg}=R_{\rm IP}R_{\rm last}^{-1},\qquad
 C_{\rm seg}=C_{\rm IP}-R_{\rm seg}C_{\rm last}.
\]
Coordinate definitions, reference momentum, matrix conditioning, and one-case
first-order propagation were checked before this construction. Direct optics
output matches the cumulative linear part to $1.92\times10^{-13}$ in the largest
coefficient.

Each particle enters the Taylor map with $(x,x',y,y',\beta_i ct_i,p_i/p_{\rm
ref}-1)$ and leaves with $p'_i=p_{\rm ref}(1+\delta'_i)$ and
$t'_i=\ell'_i/(\beta'_ic)$. The affine term includes the reference path-length
contribution. All polynomial coefficients use unique lower-triangular monomials;
no permutation multiplicity or factorial is added. Applying the second-order map
to individual particles retains the third- and fourth-order input moments that
appear when forming output covariances. A covariance-only substitution would not
provide the same calculation.

The terminal $p_{\rm ref}$ spans 383.867--400.584 in $m_ec$ units and agrees
with particle mean momentum to relative $1.56\times10^{-15}$. Maximum absolute
momentum offsets per case lie between 0.00193 and 0.00701. Native
complete-segment coefficients at reference momenta 380, 390, and 400 are
identical. The quadrupole implementation uses the lattice's normalized $K_1$,
and an independent analytic calculation confirms this lack of explicit
reference-momentum dependence. Applying the same coefficients in each case's own
reference coordinates is therefore the conditioned calculation for this
normalized-strength lattice. Rescaling $K_1$ as if a fixed physical gradient had
been specified would change that system.

Synthetic identity, affine, mixed quadratic/cubic monomials, and zero-high-order
degeneration checks passed. An isolated drift agrees with the geometric
second-order expansion. An isolated quadrupole agrees with the analytic
trigonometric/hyperbolic linear solution; its momentum derivative agrees with
second-order coefficients within $2.70\times10^{-9}$, below the $10^{-7}$
numerical tolerance. An independent truncated-polynomial composition of all 32
elements, using analytic quadrupole transfer, complex-step momentum derivatives,
and geometric path-length integration, agrees with the native full-segment
second-order coefficients to maximum absolute difference $1.14\times10^{-12}$
and relative Frobenius difference $1.13\times10^{-15}$. Neither the construction
nor these tests fit IP labels. Predictions for all 700 cases use direct
polynomial evaluation on their terminal distributions.

\subsection{Why third-order predictions are omitted}
The \textsc{elegant} release used here provides native third-order coefficients
whose lower-order blocks agree with the second-order output. An independent
quadrupole test of the quadratic-in-$\delta$ correction to the linear transverse
map gives a coefficient difference near $5.68\times10^{-4}$, exceeding the
predefined $2\times10^{-7}$ tolerance across finite-difference steps
$10^{-3}$--$10^{-5}$. This unresolved consistency discrepancy prevents
qualification of R3 as a reference, so third-order predictions are omitted. The
R2 reference is supported by the independent checks above.

\subsection{Full target results and paired margins}
\begin{table}[H]
\caption{Explicit-optics comparison on 89 unique primary OOS cases. ML uses the
frozen ensemble estimator; physics receives full particles. Stored units are
used: $\sigma_x$ and $\sigma_y$ are in mm (multiply by 1000 for the $\mu$m used in
the main text); other parameters follow the main target definitions}
\label{tab:physics-primary_OOS89}
\centering\small
\begin{tabular}{llllll}
\toprule
Target & Method & MAE & RMSE & Bias & $R^2$ \\\midrule
$\langle E_k\rangle$ & R1 & 1.28e-15 & 7.38e-15 & -6.39e-16 & 1 \\
$\langle E_k\rangle$ & R2 & 1.6e-15 & 7.97e-15 & -1.6e-15 & 1 \\
$\langle E_k\rangle$ & Terminal ML & 0.091377 & 0.16529 & -0.031467 & 0.99293 \\
$\sigma_\delta$ & R1 & 1.64e-16 & 2.34e-16 & -3.52e-17 & 1 \\
$\sigma_\delta$ & R2 & 2.19e-16 & 3.12e-16 & -7.02e-18 & 1 \\
$\sigma_\delta$ & Terminal ML & 0.0016191 & 0.002864 & 2.28e-05 & 0.99777 \\
$\sigma_x$ & R1 & 0.000972 & 0.0013461 & -0.000956 & 0.48923 \\
$\sigma_x$ & R2 & 4.17e-05 & 8.62e-05 & 2.43e-05 & 0.9979 \\
$\sigma_x$ & Terminal ML & 0.000244 & 0.000354 & -3.83e-05 & 0.96466 \\
$\sigma_y$ & R1 & 0.000498 & 0.000812 & 0.000155 & 0.24774 \\
$\sigma_y$ & R2 & 1.33e-05 & 2.36e-05 & 4.82e-06 & 0.99936 \\
$\sigma_y$ & Terminal ML & 0.00029 & 0.000529 & 1.18e-05 & 0.68075 \\
$\sigma_z$ & R1 & 0.000759 & 0.0011704 & 0.000261 & 0.99987 \\
$\sigma_z$ & R2 & 4.62e-05 & 9.65e-05 & -2.48e-06 & 1 \\
$\sigma_z$ & Terminal ML & 0.0053194 & 0.0088905 & 2.32e-05 & 0.99232 \\
$\epsilon_{nx}$ & R1 & 0.9973 & 1.6856 & -0.96727 & 0.78691 \\
$\epsilon_{nx}$ & R2 & 0.039138 & 0.091178 & 0.018491 & 0.99938 \\
$\epsilon_{nx}$ & Terminal ML & 0.27256 & 0.48545 & -0.079228 & 0.98233 \\
$\epsilon_{ny}$ & R1 & 0.69106 & 1.5176 & 0.33605 & 0.41696 \\
$\epsilon_{ny}$ & R2 & 0.01692 & 0.050092 & -0.0023949 & 0.99936 \\
$\epsilon_{ny}$ & Terminal ML & 0.33263 & 0.69414 & 0.02816 & 0.87802 \\
\bottomrule\end{tabular}
\end{table}
\begin{table}[H]
\caption{Explicit-optics comparison on external 500. ML uses the frozen ensemble
estimator; physics receives full particles. Stored units are used: $\sigma_x$
and $\sigma_y$ are in mm (multiply by 1000 for the $\mu$m used in the main
text); other parameters follow the main target definitions}
\label{tab:physics-external500}
\centering\small
\begin{tabular}{llllll}
\toprule
Target & Method & MAE & RMSE & Bias & $R^2$ \\\midrule
$\langle E_k\rangle$ & R1 & 2.33e-15 & 1.07e-14 & 2.84e-16 & 1 \\
$\langle E_k\rangle$ & R2 & 1.82e-15 & 8.81e-15 & 2.27e-16 & 1 \\
$\langle E_k\rangle$ & Terminal ML & 0.098104 & 0.15667 & -0.018267 & 0.99329 \\
$\sigma_\delta$ & R1 & 1.81e-16 & 2.46e-16 & -8.99e-18 & 1 \\
$\sigma_\delta$ & R2 & 2.31e-16 & 3.36e-16 & -8.65e-18 & 1 \\
$\sigma_\delta$ & Terminal ML & 0.0016978 & 0.002849 & 0.000225 & 0.99771 \\
$\sigma_x$ & R1 & 0.000925 & 0.0012922 & -0.000909 & 0.48431 \\
$\sigma_x$ & R2 & 3.76e-05 & 7.01e-05 & 1.72e-05 & 0.99848 \\
$\sigma_x$ & Terminal ML & 0.000249 & 0.000368 & -4.64e-06 & 0.9581 \\
$\sigma_y$ & R1 & 0.000466 & 0.00076 & 5.28e-05 & 0.52203 \\
$\sigma_y$ & R2 & 1.7e-05 & 4.24e-05 & 9.54e-07 & 0.99851 \\
$\sigma_y$ & Terminal ML & 0.00028 & 0.000425 & 1.15e-07 & 0.85 \\
$\sigma_z$ & R1 & 0.000779 & 0.0011313 & 0.000472 & 0.99988 \\
$\sigma_z$ & R2 & 4.67e-05 & 7.55e-05 & -1.29e-05 & 1 \\
$\sigma_z$ & Terminal ML & 0.0053126 & 0.0072424 & 0.000599 & 0.99447 \\
$\epsilon_{nx}$ & R1 & 0.93962 & 1.5804 & -0.90785 & 0.77416 \\
$\epsilon_{nx}$ & R2 & 0.037544 & 0.073473 & 0.01528 & 0.99951 \\
$\epsilon_{nx}$ & Terminal ML & 0.33965 & 0.58214 & 0.036534 & 0.96935 \\
$\epsilon_{ny}$ & R1 & 0.62141 & 1.2957 & 0.1493 & 0.69872 \\
$\epsilon_{ny}$ & R2 & 0.020765 & 0.056725 & -0.0064611 & 0.99942 \\
$\epsilon_{ny}$ & Terminal ML & 0.3295 & 0.58491 & 0.0035749 & 0.93845 \\
\bottomrule\end{tabular}
\end{table}
\FloatBarrier
\begin{table}[H]
\caption{Paired MAE difference, R2 minus terminal ML. Negative values favor the
full-particle R2 reference. Each target retains its stored physical unit;
$\sigma_x$ and $\sigma_y$ are in mm}
\label{tab:physics-margin}
\centering\small
\begin{tabular}{llll}
\toprule
Cohort & Target & MAE difference & 95\% interval \\\midrule
Primary OOS & $\sigma_x$ & -0.00020217 & [-0.00025473, -0.00015679] \\
Primary OOS & $\sigma_y$ & -0.00027646 & [-0.00037284, -0.00019376] \\
Primary OOS & $\epsilon_{nx}$ & -0.23342 & [-0.31341, -0.16384] \\
Primary OOS & $\epsilon_{ny}$ & -0.31571 & [-0.44657, -0.20991] \\
External 500 & $\sigma_x$ & -0.00021143 & [-0.00023337, -0.00019116] \\
External 500 & $\sigma_y$ & -0.00026295 & [-0.00028848, -0.0002389] \\
External 500 & $\epsilon_{nx}$ & -0.30211 & [-0.34123, -0.26518] \\
External 500 & $\epsilon_{ny}$ & -0.30873 & [-0.34719, -0.27296] \\
\bottomrule\end{tabular}
\end{table}
The first-order-to-second-order improvement is positive for every transverse
target in both cohorts. All eight 99.375\% intervals for R2-minus-ML MAE are
negative. These paired ensemble results favor the complete-particle R2
reference, which retains nonzero residuals. The full-state-gap table, available
from the corresponding author upon reasonable request, additionally compares R2
with A/B/C under each corresponding aggregation and common HDF5 truth.

\section{Supporting diagnostic-subset and input-uncertainty study}
\label{sec:s-vd}

The supporting diagnostic-subset study classified each stored quantity according
to whether an accelerator diagnostic can measure it directly, whether it
requires a
standard reconstruction procedure, or whether the stored simulation quantity
has no identified machine mapping. ``Directly measurable'' denotes an
identified measurement path. Reconstructible quantities can require
repeatable pulses, calibrated optics, dispersive measurements, or longitudinal
phase-space diagnostics.

Entrance-VD retains 11 of the 12 compact entrance variables. The stored mean
simulation time is excluded because its time origin has not been mapped to a
machine clock.
Six retained quantities have an identified direct measurement path and five
require reconstruction. Table~\ref{tab:s-vd-entrance-map} lists the complete
entrance classification used before fitting the reduced model.

\begin{table}[H]
\caption{Entrance-12 realizability classification defining Entrance-VD.
D denotes an identified direct measurement path, R a reconstruction path, and S
a stored simulation quantity without an identified machine mapping. The path
column describes the candidate measurement or reconstruction used in the
study}
\label{tab:s-vd-entrance-map}
\centering
\small
\begin{tabular}{@{}llll@{}}
\toprule
Stored quantity & Unit & Class & Candidate path \\
\midrule
$\langle x\rangle$ & m & D & BPM or profile centroid \\
$\sigma_x$ & m & D & calibrated transverse profile \\
$\langle x'\rangle$ & rad & R & synchronized BPMs plus transfer map \\
$\sigma_{x'}$ & rad & R & multi-profile or quadrupole-scan reconstruction \\
$\langle y\rangle$ & m & D & BPM or profile centroid \\
$\sigma_y$ & m & D & calibrated transverse profile \\
$\langle y'\rangle$ & rad & R & synchronized BPMs plus transfer map \\
$\sigma_{y'}$ & rad & R & multi-profile or quadrupole-scan reconstruction \\
$\langle t\rangle$ & s & S & no mapping for the stored time origin \\
$\sigma_t$ & s & D & calibrated longitudinal profile \\
$\langle p/(m_ec)\rangle$ & 1 & D & calibrated momentum spectrometer \\
$\sigma_{p/(m_ec)}$ & 1 & R & dispersive spectrum with corrections \\
\bottomrule
\end{tabular}
\end{table}

For Distributed-VD, 15 descriptors are retained at each of the five states and
12 entrance scalars are retained. The per-state schema is given in
Table~\ref{tab:s-vd-state-map}; the same classification is applied at all five
states. Particle count and mean simulation time are removed from every state.
The 12 retained scalar-branch entries are the 11 Entrance-VD quantities plus the
entrance relative momentum spread; particle count, mean simulation time, and the
numerical Twiss-consistency residual are removed. The resulting input contains
87 entries, of which 75 are unique state/quantity entries because several
entrance scalars duplicate entrance-state descriptors. The retained distributed
input includes quantities that can require repeatable pulses, dispersive
measurements, or longitudinal phase-space reconstruction.

\begin{table}[H]
\caption{Per-state realizability classification for the 17 M1 descriptors.
The same D/R/S definitions as in Table~\ref{tab:s-vd-entrance-map} are used.
Entries marked ``remove'' are excluded from Distributed-VD}
\label{tab:s-vd-state-map}
\centering
\small
\begin{tabular}{@{}llll@{}}
\toprule
Stored quantity & Unit & Class & Distributed-VD \\
\midrule
$N$ & count & S & remove \\
$\langle x\rangle,\sigma_x$ & m & D & retain \\
$\langle x'\rangle,\sigma_{x'}$ & rad & R & retain \\
$\operatorname{cov}(x,x')$ & m rad & R & retain \\
$\langle y\rangle,\sigma_y$ & m & D & retain \\
$\langle y'\rangle,\sigma_{y'}$ & rad & R & retain \\
$\operatorname{cov}(y,y')$ & m rad & R & retain \\
$\langle t\rangle$ & s & S & remove \\
$\sigma_t$ & s & D & retain \\
$\langle p/(m_ec)\rangle$ & 1 & D & retain \\
$\sigma_{p/(m_ec)}$ & 1 & R & retain \\
$\sigma_\delta$ & 1 & R & retain \\
$\operatorname{cov}(t,\delta)$ & s & R & retain \\
\bottomrule
\end{tabular}
\end{table}

The reduced models use the same split-42 training/validation/test assignment
(157/21/22 cases), the same five training seeds, and the same preprocessing and
hyperparameters as the corresponding full routes. Entrance-VD uses the original
64-tree ExtraTrees settings. Distributed-VD uses the M1 architecture with only
the first-layer input dimensions changed. The paired full Entrance-12 and
Entrance-VD models use the same input and target scalers, which replace
standard deviations below $10^{-12}$ by one; the reduced model selects the
corresponding input-scaler entries. Full M1 retains its primary neural scalers;
Distributed-VD uses exact-zero-only scaling for its retained moments and reuses
the retained entries of the full M1 scalar and target scalers. No feature or
hyperparameter search was performed after the diagnostic subsets were defined.

\begin{table}[H]
\caption{Performance of the full and diagnostic-subset routes. Primary and
external values are means $\pm$ sample standard deviations of the metrics
computed separately for the five trained seeds, evaluated on the same 22
held-out split-42 cases and the same separately transported 500-case cohort,
respectively}
\label{tab:s-vd-subsets}
\centering
\small
\begin{tabular}{@{}lcccc@{}}
\toprule
Route & Primary macro $R^2$ & Primary macro nMAE & 500-case macro $R^2$ &
500-case macro nMAE \\
\midrule
Full Entrance-12 & $0.9113\pm0.0060$ & $0.1809\pm0.0020$ & $0.9080\pm0.0017$ &
$0.2028\pm0.0019$ \\
Entrance-VD & $0.9153\pm0.0050$ & $0.1788\pm0.0031$ & $0.9059\pm0.0006$ &
$0.2034\pm0.0013$ \\
Full M1 & $0.8924\pm0.0316$ & $0.1325\pm0.0084$ & $0.9537\pm0.0029$ &
$0.1097\pm0.0015$ \\
Distributed-VD & $0.8252\pm0.0245$ & $0.1542\pm0.0058$ & $0.9480\pm0.0021$ &
$0.1147\pm0.0009$ \\
\bottomrule
\end{tabular}
\end{table}

\begin{figure}[H]
\centering
\includegraphics[width=0.96\linewidth]{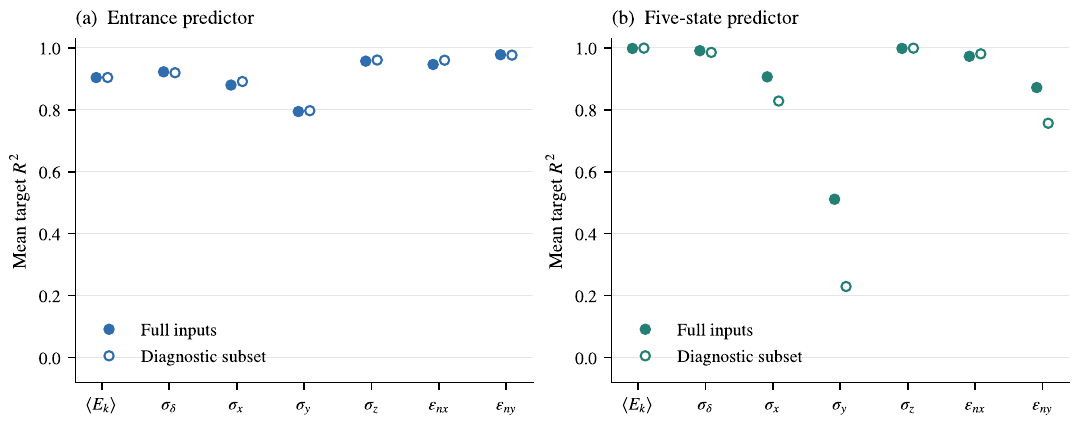}
\caption{Can a diagnostically motivated input subset retain prediction
performance? Panels compare the entrance predictor and the five-observation
predictor on the same 22 held-out simulated cases from partition 42.
Full inputs and diagnostic subsets are defined in the preceding tables.
Each point is mean $R^2$ over five training seeds; higher is better and 1 is
exact prediction. The horizontal axis lists mean energy, momentum spread,
horizontal/vertical rms size, bunch length, and horizontal/vertical normalized
emittance. Filled/open markers identify full/reduced inputs. No measurement
uncertainty is shown. This is an input-availability sensitivity test, not
validation against hardware diagnostics}
\label{fig:s-vd-subset}
\end{figure}

The primary target-level results show that Entrance-VD changes the full
Entrance-12 $R^2$ by less than 0.014 in absolute value for every target. The
largest Distributed-VD penalties occur for the transverse targets, especially
$\sigma_y$ and $\epsilon_{ny}$ on the 22-case split-42 test set. The separate
500-case cohort shows a much smaller aggregate penalty, with
$\Delta R^2=-0.0057$ relative to full M1 and an nMAE inflation of 1.046.

Input uncertainty was added only at inference time. Five levels were used:
0, 0.5, 1, 2, and 5\%. Each nonzero level uses 50 Monte Carlo realizations for
each of the five trained seeds. For a level $L$ percent, $q=L/100$ and independent standard-normal deviates $z$ are used. Positive rms scales and mean momentum are multiplied by $\exp(qz-q^2/2)$. Signed centroids, mean slopes and covariance-like quantities receive $q\,s_{\rm train}z$, with the scale computed from the training partition. Relative momentum spread follows the perturbed rms/mean momentum ratio, shared entrance channels receive the same perturbation, and covariance values are clipped to their Cauchy--Schwarz bounds. Particle count and the excluded simulation time are not perturbed. Thus ``2\%'' describes this coupled perturbation rule, rather than a uniform measurement accuracy of every diagnostic. Table~\ref{tab:s-vd-subsets} averages seven targets; Table~\ref{tab:s-vd-noise} averages the six targets excluding energy.

\begin{table}[H]
\caption{Controlled input-uncertainty sensitivity for the six beam-quality
targets excluding mean kinetic energy. Values are the mean macro $R^2$ and mean
nMAE inflation relative to the zero-added-noise diagnostic subset}
\label{tab:s-vd-noise}
\centering
\small
\begin{tabular}{@{}ccccc@{}}
\toprule
Uncertainty & \multicolumn{2}{c}{Entrance-VD} &
\multicolumn{2}{c}{Distributed-VD} \\
(\%) & macro $R^2$ & nMAE inflation & macro $R^2$ & nMAE inflation \\
\midrule
0.0 & 0.9172 & 1.000 & 0.7962 & 1.000 \\
0.5 & 0.9142 & 1.032 & 0.7946 & 1.016 \\
1.0 & 0.9083 & 1.101 & 0.7905 & 1.068 \\
2.0 & 0.8948 & 1.212 & 0.7664 & 1.282 \\
5.0 & 0.8758 & 1.357 & 0.3090 & 2.862 \\
\bottomrule
\end{tabular}
\end{table}

The Distributed-VD prediction of mean kinetic energy is more sensitive than the
six-target summary because its clean error is very small. At 0.5\%, 1\%, and
2\% controlled input uncertainty, the nMAE inflation for mean kinetic energy is
approximately 10.0, 19.9, and 39.7, respectively. These values identify the
momentum scale as a critical calibration quantity if the clean energy precision
is to be retained, within this controlled input-noise model.

With the input arrays already resident in memory, the five-model Entrance-VD
ensemble requires a median 10.10~ms for one case and Distributed-VD 1.71~ms on
the tested CPU node. These timings include normalization, inference, output
denormalization, and ensemble averaging. They exclude diagnostic acquisition,
reconstruction, and tracking.

\section{Optics-perturbation validity check}
\label{sec:s-optics-validity}

A separate validation-only selection procedure was used to define perturbations
without inspecting held-out model performance. The upstream branch uses the
first Q02DBA occurrence at $s=3.3042$~m with a prescribed $\pm8\%$ strength
change.
The downstream branch uses a coordinated final-focus knob after the last M1
observation. For that knob, focusing quadrupoles are scaled by $(1+\alpha)$ and
defocusing quadrupoles by $(1-\alpha)$, with $\alpha=\pm0.02$. The latter was
the first predeclared amplitude that passed the validation identifiability
criterion.
No model was retrained or calibrated on the perturbation data.

All 88 requested held-out runs (22 cases times four perturbation conditions)
completed with CSR disabled. For the upstream perturbation, the entrance state
remained unchanged and the four later unique M1 observation states changed.
The particle comparison also included the co-located stored final-focus-entrance
record, which is not an additional unique M1 state. The
fraction of M1 features outside their univariate training ranges was 4.32\% for
the negative perturbation and 5.91\% for the positive perturbation. For the
downstream final-focus perturbation, all entrance and M1 non-IP observation
arrays were bitwise identical to nominal while the IP truth changed.

\begin{table}[H]
\caption{Macro prediction response in the frozen optics-perturbation test. Error
inflation is the perturbed macro nMAE divided by the model's nominal macro
nMAE. For each condition, metrics are computed from the mean prediction of the
five frozen training seeds on the 22 split-42 test cases}
\label{tab:s-optics-validity}
\centering
\small
\begin{tabular}{@{}lrrrr@{}}
\toprule
Model & U$-$ & U$+$ & D$-$ & D$+$ \\
\midrule
Entrance-12 error inflation & 2.451 & 2.840 & 55.591 & 87.216 \\
M1 error inflation & 3.903 & 3.615 & 74.681 & 117.095 \\
Entrance-12 $\Delta R^2$ & $-0.3391$ & $-0.1948$ & $-1.4998$ & $-1.7219$ \\
M1 $\Delta R^2$ & $-0.7748$ & $-0.3827$ & $-1.4731$ & $-1.6274$ \\
\bottomrule
\end{tabular}
\end{table}

For the upstream perturbations, M1 has larger aggregate error inflation than
Entrance-12. The downstream perturbations identify a separate observation
boundary: identical predictor inputs correspond to different IP outcomes when
the transport map changes after the final observation. The predictor must
therefore be associated with its downstream optics configuration.

\begin{figure}[H]
\centering
\includegraphics[width=0.96\linewidth]
{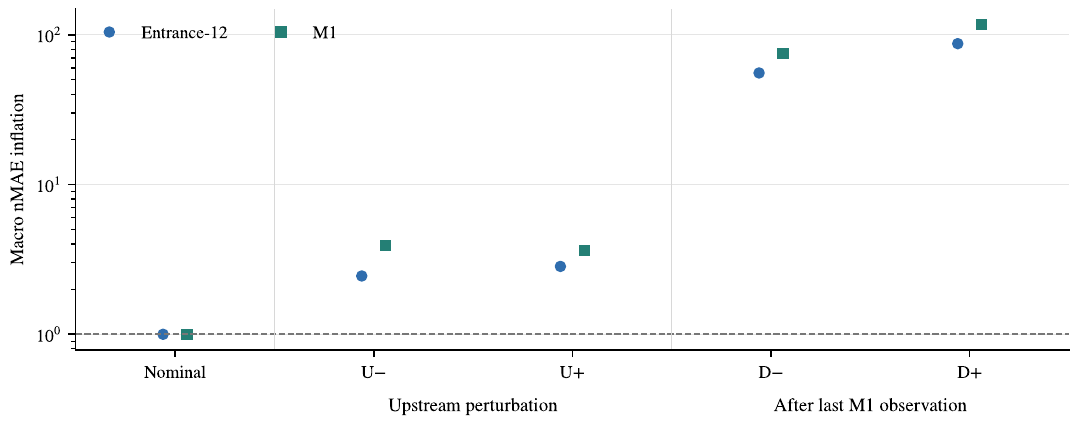}
\caption{How strongly does changing the optics affect a predictor trained
for a fixed lattice? The ordinate is error inflation: perturbed mean
normalized absolute error divided by its nominal value. Each target error
is first divided by its training-target standard deviation; the seven
normalized errors are then averaged. A value of 1 means unchanged error;
larger values mean worse prediction. The logarithmic axis shows both mild
and large changes. U$-$/U$+$ denote the negative/positive upstream Q02DBA
perturbations; D$-$/D$+$ denote coordinated final-focus perturbations after
the last observation of the five-state model (M1). The entrance and M1
models remain frozen; points use five-seed ensemble predictions for the
same 22 held-out simulated cases. The final-focus changes alter the targets
while leaving all supplied inputs unchanged, demonstrating that the
predictor must be associated with its downstream optics configuration}
\label{fig:s-optics-validity}
\end{figure}

\FloatBarrier

\section{Historical runtime boundaries}
\label{sec:s-runtime}

The earlier timing comparison was performed on 20 preselected cases with three repeated
measurements after warm-up. The Elegant reference starts from the same entrance
particle file as the Entrance-12 route. The Entrance-12 pipeline includes
entrance
feature extraction, cold loading of the predeclared split42/seed1 ExtraTrees
model, and inference.
The M1 pipeline includes
tracking to the second-achromat exit, descriptor extraction, cold model loading,
and inference.

Direct \textsc{elegant}/M1 and Entrance-12 timings were obtained in separate
jobs on different compute nodes, with each job requesting
2 CPUs and 8~GiB. The Entrance-12 job
used Python 3.13.13, NumPy 2.2.6, scikit-learn 1.7.2, and joblib 1.5.3; the
tracking executable was \textsc{elegant} 2026.2.0. The exact CPU model, actual
application thread utilization, and numerical-library thread environment were
not retained. The reported ratio compares the measured workflows across
these compute nodes.

\begin{table}[H]
\caption{Measured runtime under two input-availability conditions. The reported
ratio is $T_{\rm elegant}/T_{\rm pipeline}$, summarized as the median of
casewise ratios. The two time columns show medians over case-level timing
summaries defined below}
\label{tab:s-runtime}
\centering
\small
\begin{tabular}{@{}lccc@{}}
\toprule
Route and boundary & Elegant (s) & Pipeline (s) & Ratio \\
\midrule
Entrance-12, entrance file available & 1.170 & 0.057 & 20.55 \\
M1, second-achromat-exit state required & 1.170 & 1.252 & 0.932 \\
\bottomrule
\end{tabular}
\end{table}

For Entrance-12, each case-level time is the median over three repeats. The
paired ratio is the median of same-case, same-repeat time ratios within each
case, followed by the median over cases. For M1, each case-level pipeline time
is the sum of the separately measured tracking, descriptor-extraction, and cold
inference component medians. Its paired ratio is calculated from these
case-level summaries. For Entrance-12, the paired ratio has P10--P90 range
19.79--24.60. The first
comparison measures approximate downstream prediction from an available entrance
particle file. The M1
comparison includes the tracking needed to generate its downstream input; its
median over the 20 case-level summaries is 1.252~s and its paired speed ratio is 0.932.
Both timing boundaries start from available entrance particles and include
the input-generation costs specified above.

These values are retained as historical provenance only. They were measured on
separate jobs and do not establish the current comparison. No current main-text speed claim or figure depends on these historical timings.

\subsection{Additional historical timing limitations}
A separate archived compute-node run used the fixed 20-case utility selection, one excluded warm-up, ten repeats and rotated method order. Its file-labelled Particle-R2 interval began with resident particles and coefficients; map generation was measured separately. Elegant timing ended at the IP output file, without extracting the seven observables. ET-12 used a single split-42/seed-1 checkpoint and initially stored standardized outputs; the target scaler was subsequently used to restore physical units in the audit. These measurements do not establish a matched file-to-seven-output comparison or pair with the common-case ensemble accuracy.

The revised timing driver separates coefficient generation/loading from propagation and validation, populates resident entrance features, and uses only the frozen eligible models for each case. The fixed utility selection contains 20 cases from the additional 500-case cohort, for which all 25 models are eligible. That measured ensemble and cohort are distinct from the primary 89-case estimator, whose model count varies by held-out membership. Cold, cached-file, resident, setup and batch boundaries are recorded separately. Application-level cold loading does not imply that the operating-system file cache was cleared.

\section{Scope of the supporting CSR calculations}
\label{sec:s-csr}
The representation comparisons use radiation-free downstream labels. A separate
supporting study examined whether the available one-dimensional CSR treatment
could supply reference labels over this operating domain. Four operating
settings (cases 0001, 0002, 0007, and 0008) were evaluated with sources of
50,000 and 100,000 macroparticles at fixed 1-nC charge. The ASTRA
source-generation law was held fixed when particle count was changed; it is
distinct from the unavailable law used to sample injector controls.

Six numerical settings per case varied particle number, histogram binning,
smoothing, or post-bend kick spacing, for 24 completed tracking runs. These
supporting simulations were not used to train or evaluate the compact routes.
Qualification required charge consistency, numerical stability relative to the
CSR-effect scale, model applicability, and complete intended post-bend wake
coverage. The study's Derbenev-ratio screening criterion was $D<0.1$ at every
retained sample.

Charge consistency was satisfied, but case 0002 had maximum $D=0.1906$--0.1950
at the third bend in every numerical setting. The other three cases met this
screening criterion alone. Of 32 case/observable comparisons for numerical
sensitivity, 15 met the predefined effect-relative range criterion, five
exceeded it, and twelve lacked a resolved effect scale. The CSR-on/off values
supplied the effect scale for this test.

The implemented post-bend CSRDRIFT intervals covered only 0.2, 0.5, 0.5, and
0.6~m immediately after the four bends. Full downstream wake propagation was not
established. The combined evidence therefore did not qualify these outputs as
CSR-inclusive labels for the intended domain. The radiation-free scope follows
from this unresolved qualification, rather than from evidence that CSR is
negligible. The numerical tables are available from the corresponding author
upon reasonable request.

\section{Reproducibility and data boundaries}
Numerical metrics, paired intervals, terminal case predictions, representation
schemas, full particle files, checkpoints, and normalization statistics are
available from the corresponding author upon reasonable request.

The separately transported 500-case cohort has been used in earlier analyses.
Its joint control-sampling and screening history is incompletely documented, so
it serves as an external consistency check. The reported prediction accuracy
applies to these simulation ensembles.

\clearpage
\section{Casewise Entrance-12 prediction comparisons}
\begin{figure}[!htp]
\centering
\includegraphics[width=\linewidth]{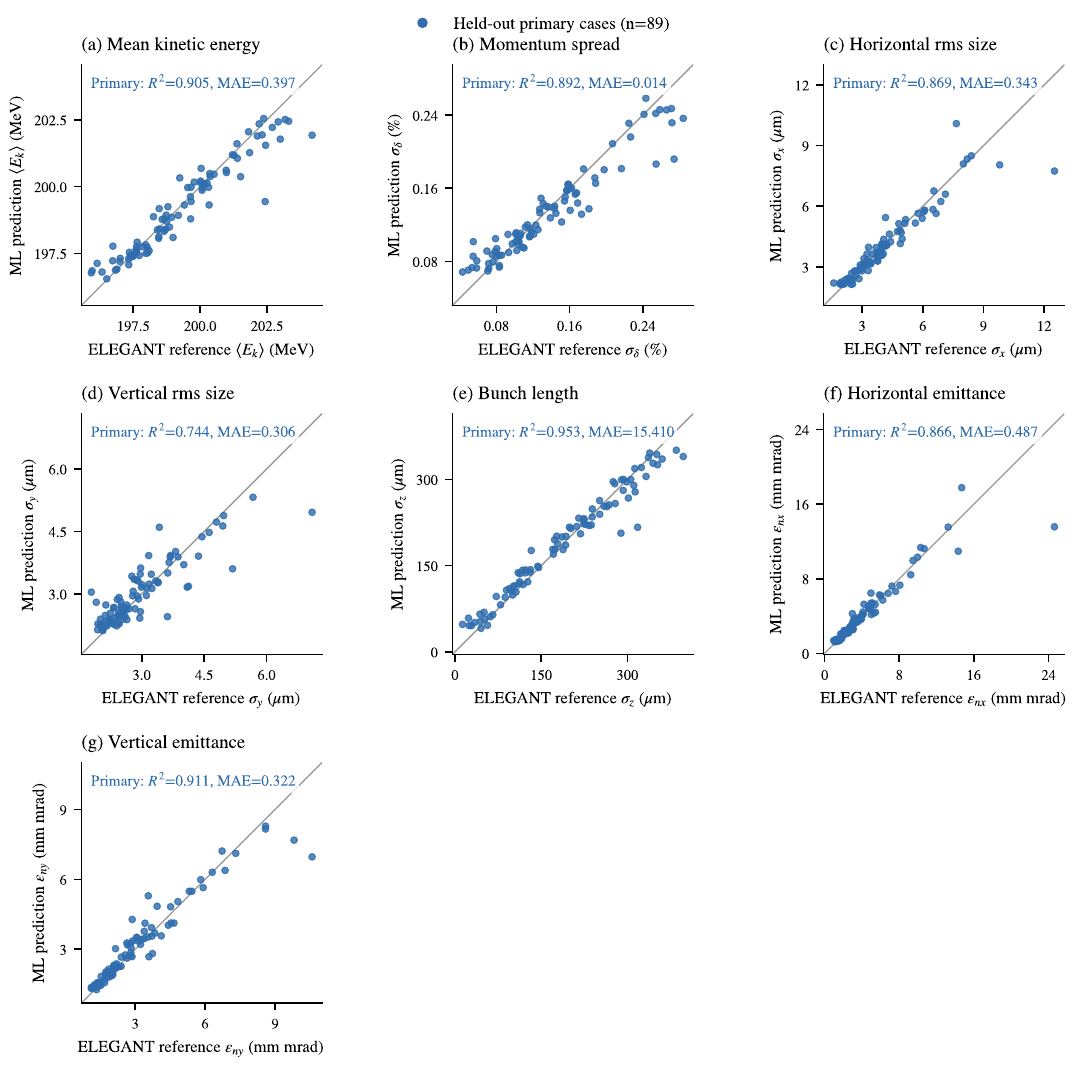}
\caption{Casewise Entrance-12 ExtraTrees predictions for all seven outputs
on the 89 distinct held-out primary cases: (a) mean kinetic energy, (b)
relative momentum spread, (c,d) horizontal and vertical rms sizes, (e) bunch
length, and (f,g) horizontal and vertical normalized emittances. Each point
uses only models for which that case belongs to the test set, with the
seed-then-partition averaging defined in the text. The diagonal denotes exact
agreement. Metrics are calculated from the plotted points; MAE uses the axis
units}
\label{fig:prediction_primary}
\end{figure}

\begin{figure}[!htp]
\centering
\includegraphics[width=\linewidth]{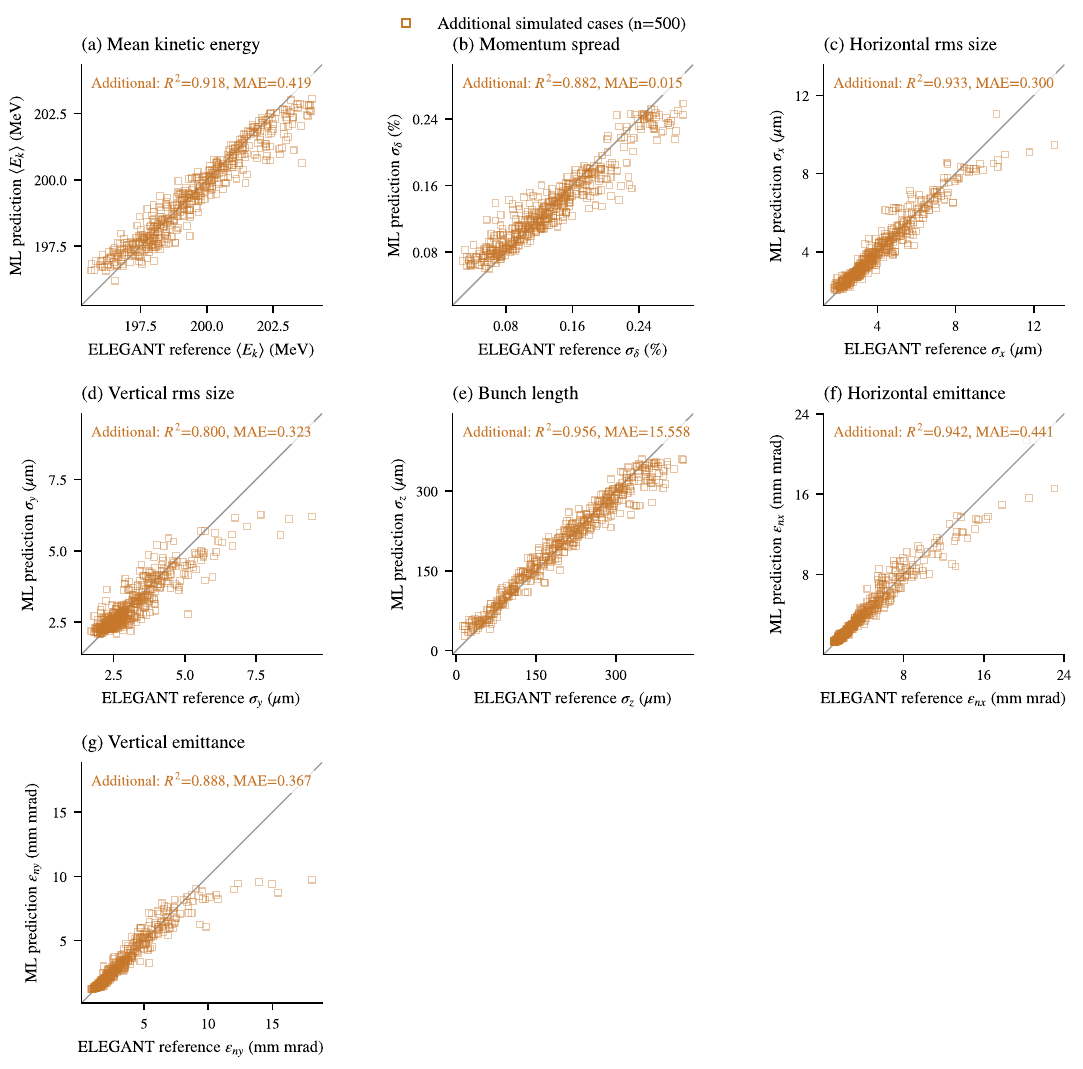}
\caption{Casewise Entrance-12 ExtraTrees predictions for all seven outputs
on the additional 500 simulated cases. Panel order and units match
Fig.~\ref{fig:prediction_primary}. Predictions average five seeds within each
of the five frozen partitions, followed by averaging of the partition
predictions. The diagonal denotes exact agreement. The displayed $R^2$ and
MAE are calculated from these plotted casewise predictions; MAE uses the axis
units}
\label{fig:prediction_additional}
\end{figure}

Figures~\ref{fig:prediction_primary} and \ref{fig:prediction_additional} show
all seven outputs of the frozen Entrance-12 ExtraTrees models. The primary
cohort contains 89 distinct cases that appear in at least one test partition.
For each case, predictions are averaged over the five training seeds within
each test partition and then over only the test partitions containing that
case. For each additional case, the same five seed predictions are averaged within
each partition and then across all five partitions. The plotted $R^2$ and MAE
are recomputed after this casewise aggregation. The historical representation comparison averages individual-model scores for the primary cohort and partition seed-ensemble scores for the external cohort, giving macro $R^2$ values of 0.881 and 0.894. The present main-text common-case comparison instead scores the 89 casewise averaged predictions and gives 0.877 for ET-12.
The 500 additional simulated cases serve as the external consistency check
described above.

\clearpage

\clearpage
\section{Entrance-map qualification and common-case reaggregation}
The entrance-map route starts from the rematched entrance particle rows.
Its coordinates are $(x,x',y,y',s,\delta)$ with $s=\beta ct$ and
$\delta=p/p_{\rm ref}-1$. The two particle cuts of the simulated lattice (RCOL, $|x|\le2$~mm; main-text Sec.~2.2) are applied in lattice order. They are not part of the design and do not model the 6-mm-inner-diameter vacuum chamber. The
entrance SDDS has no verified stable particle identifier, so stable entrance
row indices are propagated through predicted masks. IP survivors are used only
for scoring. Qualification passed coordinate conversion, reference momentum,
R/T convention, synthetic map tests, selection semantics, row-index propagation,
and a regression at 600 checkpoints comparing two implementations on the same 200 cases. Each case contributes two RCOLs and IP. The coefficient difference is zero. This test does not establish identity between different cases.

The strict paired set contains \CommonN{} cases, the intersection of the
frozen ET-12 primary OOS union, the 200-case entrance-map cohort, and the
retained truth. ET-12 uses the frozen arithmetic mean over eligible
partition/seed predictions. ET-28 and Ridge-28 are summary-only in the
retained assets and are not treated as casewise paired predictions.

\begingroup\scriptsize
\begin{longtable}{@{}lllrrrrr@{}}
\caption{Common 89-case accuracy. G denotes Gaussian closure; pp denotes percentage points. Sizes are in mm in this table (main-text plots use micrometres).}\label{tab:v23-accuracy}\\
\toprule Method & Target & Unit & $N$ & $R^2$ & MAE & RMSE & Bias\\\midrule\endfirsthead
\toprule Method & Target & Unit & $N$ & $R^2$ & MAE & RMSE & Bias\\\midrule\endhead
ET-12 & $\langle E_k\rangle$ & MeV & 89 & 0.90486 & 0.39703 & 0.60647 & -0.11825\\
ET-12 & $\epsilon_{nx}$ & mm mrad & 89 & 0.86601 & 0.48686 & 1.3366 & -0.089722\\
ET-12 & $\epsilon_{ny}$ & mm mrad & 89 & 0.91102 & 0.32166 & 0.59284 & 0.040485\\
ET-12 & $\sigma_\delta$ & pp & 89 & 0.89179 & 0.013659 & 0.019944 & -0.0035032\\
ET-12 & $\sigma_x$ & mm & 89 & 0.86878 & 0.00034273 & 0.0006823 & -9.8271e-05\\
ET-12 & $\sigma_y$ & mm & 89 & 0.74449 & 0.00030589 & 0.00047304 & 4.5499e-05\\
ET-12 & $\sigma_z$ & mm & 89 & 0.95313 & 0.01541 & 0.02197 & 0.00070544\\
Moment-R1 & $\langle E_k\rangle$ & MeV & 89 & 0.9993 & 0.015704 & 0.052201 & 0.010112\\
Moment-R1 & $\epsilon_{nx}$ & mm mrad & 89 & 0.89633 & 0.64831 & 1.1757 & -0.38592\\
Moment-R1 & $\epsilon_{ny}$ & mm mrad & 89 & 0.35194 & 0.7376 & 1.6 & 0.61352\\
Moment-R1 & $\sigma_\delta$ & pp & 89 & 0.6114 & 0.01069 & 0.037795 & 0.010686\\
Moment-R1 & $\sigma_x$ & mm & 89 & 0.54616 & 0.00087916 & 0.0012689 & -0.0008492\\
Moment-R1 & $\sigma_y$ & mm & 89 & -0.17592 & 0.00060079 & 0.0010148 & 0.00042598\\
Moment-R1 & $\sigma_z$ & mm & 89 & 0.99816 & 0.0026795 & 0.0043573 & 0.0024512\\
Moment-R2 (G) & $\langle E_k\rangle$ & MeV & 89 & 0.9993 & 0.015704 & 0.052201 & 0.010112\\
Moment-R2 (G) & $\epsilon_{nx}$ & mm mrad & 89 & -0.017237 & 1.1479 & 3.6829 & 1.0804\\
Moment-R2 (G) & $\epsilon_{ny}$ & mm mrad & 89 & -2.5481 & 1.4281 & 3.7437 & 1.4054\\
Moment-R2 (G) & $\sigma_\delta$ & pp & 89 & 0.6114 & 0.01069 & 0.037795 & 0.010686\\
Moment-R2 (G) & $\sigma_x$ & mm & 89 & 0.44692 & 0.00048207 & 0.0014008 & 0.00016015\\
Moment-R2 (G) & $\sigma_y$ & mm & 89 & -2.9826 & 0.00095376 & 0.0018676 & 0.00089675\\
Moment-R2 (G) & $\sigma_z$ & mm & 89 & 0.99797 & 0.0028345 & 0.0045701 & 0.0026653\\
Particle-R1 & $\langle E_k\rangle$ & MeV & 89 & 0.99991 & 0.004407 & 0.018949 & 0.0037491\\
Particle-R1 & $\epsilon_{nx}$ & mm mrad & 89 & 0.78058 & 1.0019 & 1.7104 & -0.97317\\
Particle-R1 & $\epsilon_{ny}$ & mm mrad & 89 & 0.42877 & 0.65494 & 1.5021 & 0.37281\\
Particle-R1 & $\sigma_\delta$ & pp & 89 & 0.99444 & 0.0010609 & 0.004521 & 0.0009824\\
Particle-R1 & $\sigma_x$ & mm & 89 & 0.47 & 0.00098079 & 0.0013712 & -0.00096507\\
Particle-R1 & $\sigma_y$ & mm & 89 & -0.0017186 & 0.00055581 & 0.00093663 & 0.00021994\\
Particle-R1 & $\sigma_z$ & mm & 89 & 0.99985 & 0.00083831 & 0.0012561 & 0.00031594\\
Particle-R2 & $\langle E_k\rangle$ & MeV & 89 & 0.99988 & 0.0049092 & 0.021281 & 0.004775\\
Particle-R2 & $\epsilon_{nx}$ & mm mrad & 89 & 0.98982 & 0.094958 & 0.36846 & 0.058348\\
Particle-R2 & $\epsilon_{ny}$ & mm mrad & 89 & 0.99012 & 0.13664 & 0.19751 & 0.13113\\
Particle-R2 & $\sigma_\delta$ & pp & 89 & 0.9932 & 0.0011238 & 0.0049995 & 0.0011234\\
Particle-R2 & $\sigma_x$ & mm & 89 & 0.98933 & 6.4428e-05 & 0.00019456 & -5.664e-06\\
Particle-R2 & $\sigma_y$ & mm & 89 & 0.97011 & 0.00013198 & 0.00016179 & 0.00012897\\
Particle-R2 & $\sigma_z$ & mm & 89 & 0.99999 & 0.00011196 & 0.00033299 & 4.349e-05\\
\bottomrule\end{longtable}
\endgroup

\begin{table}[h]
\caption{Equal-weight macro means of target-wise $R^2$ on the common cases.}
\centering
\begin{tabular}{@{}lr@{}}\toprule
Method & Macro $R^2$\\\midrule
ET-12 & \CommonETMacroRtwo{}\\
Moment-R1 & \CommonMomentRoneMacroRtwo{}\\
Moment-R2 (G) & \CommonMomentRtwoMacroRtwo{}\\
Particle-R1 & \CommonParticleRoneMacroRtwo{}\\
Particle-R2 & \CommonParticleRtwoMacroRtwo{}\\
\bottomrule\end{tabular}
\end{table}

Paired error differences are defined as $|e_A|-|e_B|$; negative values favor the left method. The complete results are given in Table~\ref{tab:v23-paired}.
\begingroup\scriptsize
\begin{longtable}{@{}lllrr@{}}
\caption{Paired absolute-error differences, left minus right, on 89 cases; negative favors left. Intervals use 10,000 case-bootstrap replicates. Units match Table~\ref{tab:v23-accuracy}.}\label{tab:v23-paired}\\
\toprule Left & Right & Target & Mean difference & 95\% interval\\\midrule\endfirsthead
\toprule Left & Right & Target & Mean difference & 95\% interval\\\midrule\endhead
ET-12 & Particle-R2 & $\langle E_k\rangle$ & 0.39212 & [0.30318, 0.49692]\\
ET-12 & Particle-R2 & $\epsilon_{nx}$ & 0.3919 & [0.2329, 0.62971]\\
ET-12 & Particle-R2 & $\epsilon_{ny}$ & 0.18502 & [0.10331, 0.28834]\\
ET-12 & Particle-R2 & $\sigma_\delta$ & 0.012536 & [0.0097343, 0.015744]\\
ET-12 & Particle-R2 & $\sigma_x$ & 0.00027831 & [0.00018601, 0.00039848]\\
ET-12 & Particle-R2 & $\sigma_y$ & 0.00017391 & [0.00011093, 0.00024366]\\
ET-12 & Particle-R2 & $\sigma_z$ & 0.015298 & [0.012303, 0.018816]\\
Moment-R2 (G) & Particle-R2 & $\langle E_k\rangle$ & 0.010794 & [0.0052404, 0.017921]\\
Moment-R2 (G) & Particle-R2 & $\epsilon_{nx}$ & 1.053 & [0.51366, 1.8142]\\
Moment-R2 (G) & Particle-R2 & $\epsilon_{ny}$ & 1.2915 & [0.67693, 2.0645]\\
Moment-R2 (G) & Particle-R2 & $\sigma_\delta$ & 0.0095657 & [0.0036336, 0.017764]\\
Moment-R2 (G) & Particle-R2 & $\sigma_x$ & 0.00041764 & [0.00022143, 0.00071222]\\
Moment-R2 (G) & Particle-R2 & $\sigma_y$ & 0.00082178 & [0.00052781, 0.0011814]\\
Moment-R2 (G) & Particle-R2 & $\sigma_z$ & 0.0027225 & [0.0020373, 0.0034954]\\
Moment-R1 & Moment-R2 (G) & $\langle E_k\rangle$ & 0 & [0, 0]\\
Moment-R1 & Moment-R2 (G) & $\epsilon_{nx}$ & -0.49964 & [-1.2968, 0.044945]\\
Moment-R1 & Moment-R2 (G) & $\epsilon_{ny}$ & -0.69052 & [-1.2009, -0.30939]\\
Moment-R1 & Moment-R2 (G) & $\sigma_\delta$ & 0 & [0, 0]\\
Moment-R1 & Moment-R2 (G) & $\sigma_x$ & 0.00039708 & [6.2682e-05, 0.00066129]\\
Moment-R1 & Moment-R2 (G) & $\sigma_y$ & -0.00035297 & [-0.00056933, -0.00017774]\\
Moment-R1 & Moment-R2 (G) & $\sigma_z$ & -0.00015497 & [-0.00032541, -2.5283e-05]\\
\bottomrule\end{longtable}
\endgroup

\section{Injector settings versus entrance beam moments on the common cases}
\label{sec:s-settings}
The settings model uses the same ExtraTrees configuration as the entrance-moment
model with the eight injector controls as input. Its retained held-out
predictions (one per partition) are averaged for each of the 89 common cases,
and the entrance-moment predictions are averaged over all held-out partition and
seed models, as in the main text. Both are scored against the same Elegant
truth. The original partition-averaged macro $R^2$ values (0.548 for settings,
0.881 for entrance moments) use a different aggregation and are reported
separately in the main text.

\begin{table}[htbp]
\caption{ExtraTrees with injector settings or entrance means and rms widths as input, 89 common held-out cases}
\label{tab:s-settings}
\centering\small
\begin{tabular}{@{}lcccc@{}}\toprule
& \multicolumn{2}{c}{Injector settings} & \multicolumn{2}{c}{Entrance moments} \\
\cmidrule(lr){2-3}\cmidrule(l){4-5}
Parameter & $R^2$ & MAE & $R^2$ & MAE \\\midrule
$\langle E_k\rangle$ (MeV) & 0.940 & 0.349 & 0.905 & 0.397 \\
$\sigma_\delta$ (\%) & 0.618 & 0.0296 & 0.892 & 0.0137 \\
$\sigma_x$ ($\mu$m) & 0.525 & 0.979 & 0.869 & 0.343 \\
$\sigma_y$ ($\mu$m) & 0.218 & 0.643 & 0.744 & 0.306 \\
$\sigma_z$ (mm) & 0.927 & 0.0212 & 0.953 & 0.0154 \\
$\epsilon_{nx}$ (mm\,mrad) & 0.521 & 1.68 & 0.866 & 0.487 \\
$\epsilon_{ny}$ (mm\,mrad) & 0.343 & 1.22 & 0.911 & 0.322 \\
\midrule
Average $R^2$ & 0.585 & & 0.877 & \\
\bottomrule\end{tabular}
\end{table}

\section{Map checks and second-order Gaussian moment propagation}
The six cases with terminal transmission below 0.95 account for 91.1\% of the ET-12 horizontal-emittance squared error and 79.1\% of the horizontal-size squared error. These terminal transmission strata are descriptive and do not select inputs or fitted models. The three-stratum and two-stratum tables use each stratum's own true-target mean in the $R^2$ denominator; the six-case values are not used to make performance claims. On the remaining 83 cases, horizontal-emittance $R^2$ is 0.96725.

Because the labels contain the two particle cuts, the map predictor must remove the same particles to reproduce them. Switching the cut test off in the second-order map, with everything else unchanged, increases the mean absolute error on the 89 cases from 0.064 to 0.965~$\mu$m for $\sigma_x$, from 0.132 to 0.337~$\mu$m for $\sigma_y$, from 0.095 to 1.758~mm\,mrad for $\epsilon_{nx}$ and from 0.137 to 0.531~mm\,mrad for $\epsilon_{ny}$. This measures how strongly the cuts of the simulated lattice affect the transverse labels; it is not a property of the design, which does not contain these cuts.

The deterministic polynomial check directly imports the production moment-propagation function. All 28 individual constant, linear and unique quadratic terms, including cross terms, are checked, followed by a random mixture. Independent three-node-per-coordinate Gauss--Hermite quadrature integrates the degree-four polynomials needed for their output covariance exactly. The largest residual is $1.42\times10^{-14}$ against a $10^{-10}$ algebraic tolerance. This tests production coefficient conversion and propagation, rather than a duplicate of the same formula. It neither proves that real beam distributions are Gaussian nor excludes defects outside the tested path.

Recovered coefficient arrays have shape $[200,3,6,28]$: cases, checkpoints, output coordinates and monomials. At both RCOLs ($s=3.3042136$ and 19.9137004~m) and IP ($s=32.6252140$~m), all 200 primary cases have identical coefficient values and the same lattice hash. They have 200 different reference momenta. Thus coefficient identity holds for these cases, while $s=\beta(p)ct$, $\delta=p/p_{\rm ref}-1$ and the inverse coordinate conversion remain case specific. Timing retains per-case map generation and cache identity; no general reuse claim is made for other configurations or the separate timing cohort.

The no-selection Physics-12 baseline uses only the stored entrance mean and rms momentum. Its energy estimate is $f(\langle p\rangle)=(\sqrt{1+\langle p\rangle^2}-1)m_ec^2$, whereas particle truth is $\langle f(p)\rangle$; these are not exactly equal. Its relative momentum-spread estimate is $100\sigma_p/\langle p\rangle$. Table~\ref{tab:v23-physics12} uses the same 89 cases and truth as the main comparison. Differences from Moment-R1 are consistent with the stored float32 entrance descriptors: the maximum energy discrepancy is $7.74\times10^{-6}$~MeV and the maximum momentum-spread discrepancy is $3.46\times10^{-8}$ percentage points. No covariance or tracking-survivor information is supplied to this baseline. Bunch length and the four transverse outputs are not determined by these 12 descriptors alone: the required cross-coordinate covariances and particle-selection information are absent.
\begingroup\scriptsize
\begin{longtable}{@{}lrrrr@{}}
\caption{No-selection Physics-12 baseline from entrance mean and rms momentum only. Energy uses $f(\langle p\rangle)$, not $\langle f(p)\rangle$. Units match Table~\ref{tab:v23-accuracy}. The last column gives the largest prediction difference from Moment-R1.}\label{tab:v23-physics12}\\
\toprule Target & $N$ & $R^2$ & MAE & Max. difference\\\midrule\endfirsthead
\toprule Target & $N$ & $R^2$ & MAE & Max. difference\\\midrule\endhead
$\langle E_k\rangle$ & 89 & 0.999295 & 0.0157036 & 7.74e-06\\
$\sigma_\delta$ & 89 & 0.611397 & 0.0106895 & 3.46e-08\\
\bottomrule\end{longtable}
\endgroup

\section{Same-node execution-cost comparison}
A fixed set of 20 cases from the additional 500-case cohort was selected by the pre-existing survival-stratified rule, independently of prediction error. The completed allocation ran on one Intel Xeon Platinum 8358P node at 2.60~GHz, with two allocated CPUs and 8~GiB memory. OMP, MKL, OpenBLAS and NumExpr thread settings were all one. Software versions were Python 3.13.13 and scikit-learn 1.7.2; the complete environment and executable/model hashes accompany the reproducibility records. A warm-up for each applicable method/boundary was excluded, followed by ten repeats with rotated method order. The 20-case batch also had one excluded full warm-up per method.

Cold calls include entrance-file parsing and all seven outputs. ET-12 loads the 25 eligible models and applies their original input and output scalers; Particle-R2 generates and parses the cumulative maps before propagation and predicted selection; Elegant stages the frozen case, tracks and extracts the seven observables. Cached-file ET retains models, while cached-file Particle-R2 reads a previously generated map; both still read the entrance particles. Resident ET has features and models in memory, and resident Particle-R2 has map coordinates and coefficients in memory. Resident Elegant still launches a fresh process from staged files. The setup interval contains model loading or coefficient generation/parsing only, excluding validation and particle propagation. No operating-system page cache was cleared.

\begingroup\small
\begin{longtable}{@{}llrr@{}}
\caption{Same-node wall times on the fixed 20-case external-cohort selection. All times are seconds; batch rows are totals for 20 cases. Intervals are the interquartile range of 200 single-case measurements or 10 full batches. ET-12 uses 25 eligible frozen models.}\label{tab:v23-timing}\\
\toprule Boundary & Method & Median & Interquartile range\\\midrule\endfirsthead
\toprule Boundary & Method & Median & Interquartile range\\\midrule\endhead
Batch (20 cases) & Elegant & 27.596 & [27.519, 27.755]\\
Batch (20 cases) & ET-12 (25 models) & 2.2177 & [2.206, 2.233]\\
Batch (20 cases) & Particle-R2 & 2.7471 & [2.7083, 2.7824]\\
Cached file & ET-12 (25 models) & 0.10226 & [0.098712, 0.10506]\\
Cached file & Particle-R2 & 0.093273 & [0.088727, 0.099]\\
Cold file & Elegant & 1.3899 & [1.3749, 1.4116]\\
Cold file & ET-12 (25 models) & 0.33419 & [0.33075, 0.33891]\\
Cold file & Particle-R2 & 0.13823 & [0.13546, 0.14255]\\
Resident & Elegant & 1.3666 & [1.3552, 1.3833]\\
Resident & ET-12 (25 models) & 0.050831 & [0.050667, 0.052747]\\
Resident & Particle-R2 & 0.018268 & [0.017906, 0.019541]\\
Setup & ET-12 (25 models) & 0.2351 & [0.2318, 0.2365]\\
Setup & Particle-R2 & 0.068886 & [0.066115, 0.073426]\\
\bottomrule\end{longtable}
\endgroup

Single-case summaries pool 20 cases times ten repeats. The paired ratio first takes Elegant/route for each case and repeat, then the median within each case, then the median across cases. Complete-call paired ratios are 4.157 for ET-12 and 10.046 for Particle-R2; resident ratios are 26.881 and 74.956. Batch ratios pair full batches by repeat and are 12.479 and 10.068. The ratio of method medians is retained separately and is not substituted for the paired statistic. Full batches include a single load of the unique 25-model set for ET-12 and a fresh map for every Particle-R2 case.

There are 2,030 measured timing records and 420 case--method--target accuracy records. Predictions have zero observed variation across repeats and available single-case boundaries; the new Elegant outputs have zero observed difference from the retained reference. Model, selection and map-source hashes match the frozen inputs. The ET predictions agree with the retained external-case ensemble to floating-point roundoff (maximum energy difference $2.84\times10^{-14}$~MeV). Maps were generated within the allocation and cached only for the same case. Each cache identity records reference momentum, lattice, map source, executable, coordinate convention, order and all three cumulative checkpoints. These timing results are separate from the 89-case primary comparison and do not replace its accuracy evidence.

\section{Scope boundaries}
The additional 500 cases are a consistency cohort, not a fresh blind test.
Entrance-VD is a candidate realizability study, not an online diagnostic
system. The qualified downstream truth is radiation-free and does not qualify
CSR-inclusive labels. The lattice uses frozen normalized strengths; fixed-field
deployment remains a gap. No terminal last-WATCH map is described as an
entrance-to-IP map.